\documentclass[nofootinbib,aps,11pt,preprintnumbers,unsortedaddress,prd]{revtex4}
\usepackage{graphicx}
\usepackage{cancel}
\usepackage{amssymb}
\usepackage{textcomp}
\usepackage{amsmath}
\usepackage{bm}
\usepackage{times}
\usepackage{epsfig}
\usepackage{color}
\usepackage{graphics}
\usepackage{hyperref}
\usepackage{setspace}
\usepackage{comment}

\hypersetup{
    pdfnewwindow=true,      
    colorlinks=true,       
    linkcolor=black,          
    citecolor=blue,        
    filecolor=blue,      
    urlcolor=blue           
}

\usepackage[utf8]{inputenc}
\usepackage[ngerman, english]{babel}
\usepackage{textcomp}
\usepackage{tikz-feynman}
\usepackage{braket}
\usepackage{siunitx}
\usepackage{mathrsfs}
\usepackage{longtable}
\usepackage[a4paper, left=2.5cm, right=2.5cm, top=2.5cm]{geometry}
\numberwithin{equation}{section}

\begin{document}

\title{\Large {\it{\bf{Minimal Radiative Neutrino Masses}}}}
\author{Christiane Klein}
\author{Manfred Lindner}
\author{Sebastian Ohmer}
\affiliation{\vspace{0.15cm} \\  Particle and Astro-Particle Physics Division \\
Max-Planck-Institut für Kernphysik {\rm{(MPIK)}} \\
Saupfercheckweg 1, 69117 Heidelberg, Germany}
\begin{abstract}
We conduct a systematic search for neutrino mass models which only radiatively produce the dimension-5 Weinberg operator. We thereby do not allow for additional symmetries beyond the Standard Model gauge symmetry and we restrict ourselves to minimal models. We also include stable fractionally charged and coloured particles in our search. Additionally, we proof that there is a unique model with three new fermionic representations where no new scalars are required to generate neutrino masses at loop level. This model further has a potential dark matter candidate and introduces a general mechanism for loop-suppression of the neutrino mass via a fermionic ladder.
\end{abstract}
\maketitle

\section{Introduction}
\label{sec:intro}

Since the discovery of neutrino oscillations and, therefore, non-vanishing neutrino masses, there have been numerous proposals explaining these small but non-zero masses. All of them add at least one new representation to the Standard Model of particle physics (SM). The Standard Model is a gauge theory based on the symmetry group
\begin{align}
G_\text{SM} = SU(3)_C \times SU(2)_L \times U(1)_Y \,,
\label{eq:SMgroup}
\end{align}
and contains the scalars and fermions shown in table~(\ref{tab:1}).

\begin{longtable}{|c|c|c|}
\hline
 Name & Label & Representation\\ \hline
 Left-handed lepton doublet & $\ell_L$ & (1,2,-1) \\ \hline
 Right-handed charged fermion & $e_R$ & (1,1,-2) \\ \hline
 Left-handed quark doublet & $Q_L$ & (3,2,1/3) \\ \hline
 Right-handed up-quark & $u_R$ & (3,1,4/3) \\ \hline
 Right-handed down-quark & $d_R$ & (3,1,-2/3) \\ \hline
 Higgs boson & $H$ & (1,2,1) \\ \hline 
 \caption{The Standard Model fermion and scalar content.}
 \label{tab:1}
\end{longtable}
Here, hypercharge is normalised such that the electric charge $Q_\text{em}$ is given by $Q_\text{em} = Y/2+I_3$ with $I_3$ the third component of the weak isospin.\\

The most minimalistic extensions of the Standard Model which include neutrino masses are the well-studied seesaw mechanisms. They each add one new representation to the Standard Model.\\
Type I seesaw adds a fermionic singlet with respect to the Standard Model gauge symmetry, $\nu_R \sim (1,1,0)$\cite{Minkowski1977, Mohapatra1979, Yanagida1979, GellMann1980}. This introduces the following interactions
\begin{align}
\mathcal{L} \supset y_\ell \bar{\ell}_L \tilde{H} \nu_R + M_{\nu_R} \overline{\nu_R^c} \nu_R \,,
\end{align}
with $\tilde{H} = i \sigma_2 H^*$. The type II seesaw enlarges the Standard Model scalar content by a scalar $SU(2)_L$-triplet, $\Delta \sim (1, 3, 2)$ ~\cite{Schechter1980, Schechter1981, Lazarides1980, Mohapatra1980}, yielding the following interactions
\begin{align}
\mathcal{L} \supset y_\Delta \ell_L \Delta \ell_L + h H \Delta^\dagger H \,.
\end{align}
In type III seesaw, a fermionic $SU(2)_L$-triplet, $\rho_R \sim (1,3,0)$, is added to the Standard Model ~\cite{Foot1988}, which leads to the interactions
\begin{align}
\mathcal{L} \supset y_\rho \bar{\ell}_L \tilde{H} \rho_R + M_{\rho_R} \overline{\rho_R^c} \rho_R\,.
\end{align}

The essence of the seesaw mechanisms thereby is the suppression of the neutrino mass due to the very heavy new representations. However, besides being difficult to experimentally test, these new heavy representation can contribute to the quantum corrections of the Higgs mass. A tuning would be necessary to explain the measured Higgs mass, see Ref.~\cite{Vissani1997, Casas2004, Abada2007, Farina2013, Clarke2015, Fabbrichesi2015, Clarke2015b, Chabab2015, Clarke2016, Salvio2016, Bambhaniya2016, Dev2017} for recent discussions. Moreover, none of the type I, II, and III seesaw fields has a clear theoretical motivation. Therefore, it is appealing to think about the possibility that the neutrino masses are pure quantum effects, or, in other words, radiatively generated by loop effects. An additional loop suppression of the neutrino mass could allow for smaller masses of the new particles due to the additional loop suppression. This could avoid the neutrino hierarchy problem, while, at the same time, allowing for better testability.

The first higher order operator generating neutrino masses is the dimension-5 Weinberg operator
\begin{align}
\label{eq:Weinberg}
\mathcal{O}_W = c_W \frac{\ell_L H H \ell_L}{\Lambda_L}\, .
\end{align}
Whereas, the discussed seesaw mechanisms are tree-level realisations of the dimension-5 Weinberg operator, we will focus on models with additional suppression. This includes generation of the dimension-5 Weinberg operator at loop-level or suppression by a higher dimension of the operators inducing neutrino masses at tree level. Since the latter also induce the Weinberg operator at some loop level, we will refer to these models as radiative neutrino mass models. Known radiative neutrino mass models are for example the Zee-model~\cite{Zee1980}, the Zee-Babu model~\cite{Zee1985, Babu1988}, and the colored seesaw models~\cite{Perez2009}. Moreover, models like the scotogenic model~\cite{Ma2006} do not only introduce radiative neutrino masses but also provide a candidate for particle dark matter.

An effort has been made to find and understand radiative neutrino mass models systematically. This task was thereby approached from different angles. For example in Ref.~\cite{Babu2001, Angel2012} an effective field theory approach to organize neutrino mass mechanisms is used. Whereas, Ref.~\cite{Bonnet2012, Sierra2014, Cai2014, Cai2017, Cepedello2017} categorise neutrino mass mechanisms by the topology of the diagram which generates non-vanishing masses. Additional systematic studies of neutrino mass generation can be found in Ref.~\cite{Ma2009, Ma1998, Sierra2015}. 

In this paper, we conduct a systematic scan to answer the question: What are the neutrino mass mechanisms which generate the dimension-5 Weinberg operator at loop level and require the least number of additional Standard Model representations?  Note that we thereby have to differentiate new fermionic and scalar fields. For example, in order to generate two non-vanishing neutrino mass differences at least two new fermionic representations in the seesaw type I and III scenario have to be added, whereas only one new scalar field in the seesaw type II scenario is necessary. We therefore first focus on the minimal number of new representations giving rise to a single massive neutrino generation and later discuss additional copies when necessary for proper mixing. Additionally, single new fermionic representations have to be added as vector-like Dirac particles or Majorana fermions with zero hypercharge to avoid anomalies\footnote{Note that if multiple fermionic representations are added to the Standard Model, this no longer has to be the case.}. We will always explicitly state which combination of fermion fields we add.

In addition to pure radiative models, we will also find models introducing higher-dimensional versions of the Weinberg operator of the form 
\begin{equation}
\mathcal{O'}_W =  \frac{\ell_L H H \ell_L}{\Lambda_L^{1+2n}}(H^\dagger H)^n \, \, .
\end{equation}
However, these operators always also induce  the dimension-5 Weinberg operator, by connecting the $H^\dagger H$-pairs via loops. A rough estimate shows that the loop-suppressed dimension-5 operator will give a larger contribution than the higher dimensional tree-level operator if 
\begin{align}
\Lambda \gtrsim 4 \pi v \, \, ,
\end{align}
where $v$ is the Higgs vacuum expectation value and the scale $\Lambda$ is associated with the mass of the new particles. If there are different couplings contributing to the different operators, their ratio will also appear in this estimate \cite{Anamiati2018}.

Apart from the seesaw mechanisms, there is exactly one model that produces neutrino masses at one-loop level with only one new beyond the Standard Model representation and a second copy of the Standard Model Higgs. This is the known Zee-model~\cite{Zee1980}. Going beyond this minimal model, we systematically search for radiative neutrino mass mechanisms with two new fields. A similar approach was discussed in Ref.~\cite{Law2013}. However, in this paper we will focus on the number of representations and will allow for higher $SU(2)_L$-representations and new coloured fields.

Additionally, we will proof that radiative neutrino masses cannot be generated with only two new fermionic representations and we will further present the unique mechanism to generate neutrino masses at loop-level with just three new fermionic representations without any new scalar fields. We will also comment on the possibility of dark matter in this scenario and on the possibility to generate large loop-suppression for the neutrino mass.

The paper is organized as follows. In section~\ref{sec:models}, we will describe the systematic search for all possible radiative neutrino mass models with two new beyond the Standard Model representations. The discussion is thereby split into new fields transforming trivially with respect to $SU(3)_C$ and fields carrying colour. We continue in section~\ref{sec:minFerm} with a formal proof why there is no model generating neutrino masses at the quantum level with just two new fermions and will then introduce the unique model with three new fermionic representations. Finally, we conclude in section~\ref{sec:conc}.

\section{Search for radiative neutrino mass models}
\label{sec:models}

In this section, we describe our systematic search for radiative neutrino mass models. We employ the following set of assumptions:
\begin{itemize}
\item The Weinberg operator~(\ref{eq:Weinberg}) appears only at loop-level.
\item There are no new symmetries beyond the Standard Model symmetry group (\ref{eq:SMgroup}).
\item There is only a minimal number of new $G_\text{SM}$ multiplets involved in the mass generation for a single neutrino generation.
\end{itemize}
The first point leads to the exclusion of the type I, II, and III seesaw fields, as discussed in the introduction. The last two constraints are implemented to allow for a bottom-up search for minimal viable models.

To identify viable models we pursue the following line of arguments. If we do not allow for new symmetries, the accidental global symmetry of lepton number is broken by the neutrino mass mechanism. The new fields must therefore induce lepton number violation (LNV) by two units \footnote{To generate Dirac neutrino masses, additional symmetries are required. $U(1)_{B-L}$ is suitable for example and is discussed in Ref.~\cite{Bonilla2018, Calle2018, Chulia2018, Chulia2018a}}. However, LNV is only introduced if at least one of the new fields has a coupling to Standard Model leptons. Hence, we should be able to identify all possible candidates by systematically scanning all possible couplings to the Standard Model leptons.

After identifying a set of candidates, we check all possible interactions of the candidates to Standard Model fields and to each other. If it is not possible to choose unique and non-trivial lepton numbers for the new fields and we find LNV by two units, we have identified a possible model. This is a generalisation of the LNV argument in~\cite{Law2013}.

When conducting a systematic model scan where minimality is an important criteria, it is essential to define minimality properly. We consider models as minimal, if the number of new representations required to generate radiative neutrino masses for a single generation is minimal. We thereby differentiate three types of new fields:
\begin{itemize}
\item new scalar representations,
\item new Dirac fermions,
\item new Majorana fermions.
\end{itemize}
When counting the number of new representations, we do not consider multiple copies of the same representation and we count vector-like Dirac fermion pairs as a single new representation. The minimal number of new representations which we have to add to the Standard Model to generate neutrino masses at the quantum level is two, except for the already mentioned Zee-model.

In the following, we differentiate type A models with two new scalars, type B models with one new scalar and one new vector-like Dirac fermion, and type C models with one new scalar and one new Majorana fermion. We summarize our results in tables where we use the following terminology:
\begin{itemize}
\item Within any type, the models are ordered by the dimension of the highest $SU(2)_L$-representation of the new fields. If this coincides for two models, the representation of the second field and the hypercharges are considered.  For coloured fields, we order first by highest $SU(3)_C$-representation and then by highest $SU(2)_L$-representation.  
\item If the new scalars are colour singlets, they are labelled $\phi_{(a,b)}$, where $a$ denotes the $SU(2)_L$-representation and $b$ the hypercharge. If the new scalars transform non-trivially with respect to $SU(3)_C$, we label them $\phi_{(c,a,b)}$, where $c$ is the $SU(3)_C$-representation.
\item If the new fermions are colour singlets, they are denoted by $\psi^{(a,b)}_X$, where $a$ is the $SU(2)_L$-representation, and $b$ is the hypercharge, and X can either be L or R for a left- or right-handed fermion. If a vector-like Dirac fermion is required both chiralities are noted separately for clarity. If the new fermions carry colour, they are labelled $\psi^{(c,a,b)}_X$, respectively.
\end{itemize}

\subsection{Minimal radiative neutrino mass models without colour}

We start our systematic search by considering new representations without colour. Hence, we have to consider the leptonic fermion bilinears of the Standard Model and the possible couplings of a Standard Model lepton to a scalar field. The complete lists of minimal models of type A, B, and C without new coloured representations found in scanning these interactions are presented in the next two subsections.

\subsubsection{Minimal models with two new scalars without colour}

We present all models with up to two new uncoloured scalar representations added to the Standard Model to generate neutrino masses at the quantum level in Table~\ref{tab:A}.

The first model (A0) on the list is the Zee-model~\cite{Zee1980}, which was already mentioned. In addition to the new scalar singlet it requires a second Higgs field. Antisymmetry of the mass matrix in the flavour indices can be avoided if both Higgs doublets develop non-zero vacuum expectation values. This can also lead to flavour-violating effects~\cite{Law2013}.

\begin{longtable}{|c|p{3cm}|c|c|c|c|p{3cm}|}
\hline
Model & New fields & Loops & Eff. dim. & Relevant interactions & New? & Comments \\ \hline \endhead
A0 & $\phi_{(1,2)}$ & 1& 5 &\parbox[c]{3.7cm}{ $y_1\phi_{(1,2)}\overline{\ell_L^c} \ell_L $\\$+ \mu H_2^{\dagger}\tilde{H_1}\phi_{(1,2)}$\\$+y_2\overline{\ell_L}H_1e_R$} & no & \parbox[c]{3cm}{Zee-model~\cite{Zee1980};\\ requires two Higgs\\ doublets}\\ \hline
A1 & $\phi_{(1,2)}$, $\phi_{(1,4)}$ & 2 & 5 &\parbox[c]{3.7cm}{ $\mu \phi_{(1,4)}^{\dagger}\phi_{(1,2)}\phi_{(1,2)}$\\$+y_1 \overline{\ell_L^c} \phi_{(1,2)} \ell_L$\\$ + y_2\overline{e_R^c}\phi_{(1,4)}e_R$} & no & Zee-Babu model~\cite{Zee1985, Babu1988}\\ \hline
A2 & $\phi_{(1,2)}$, $\phi_{(2,3)}$ & 2 & 5 &\parbox[c]{3.7cm}{ $\mu \phi_{(1,2)}^{\dagger}H^{\dagger}\phi_{(2,3)}$\\ $+\lambda \tilde{H}^{\dagger}\phi_{(2,3)}\tilde{\phi}_{(1,2)}\tilde{\phi}_{(1,2)}$\\ $+y\overline{\ell_L^c} \phi_{(1,2)} \ell_L $\\$+ \frac{1}{2}(D_\mu \phi_{(2,3)})^\dagger(D^\mu \phi_{(2,3)})$} & no & \parbox[c]{3cm}{discussed in~\cite{Law2013};\\no proper mixing;\\ruled out} \\ \hline
A3 & $\phi_{(1,4)}$, $\phi_{(2,3)}$ & 2 & 5 &\parbox[c]{3.7cm}{ $\mu \phi_{(2,3)}^{\dagger}\tilde{H} \phi_{(1,4)}$\\ $+\lambda H^{\dagger}\phi_{(2,3)}H^\dagger \tilde{H}$\\ $+y\overline{e_R^c} \phi_{(1,4)} e_R $} & no & \parbox[c]{3cm}{discussed in~\cite{Law2013};\\ requires two Higgs\\ doublets} \\ \hline
A4 & $\phi_{(1,2)}$, $\phi_{(3,0)}$ & 1 & 7 & \parbox[c]{3.7cm}{$\mu(H^{\dagger}\sigma_a H)\phi_{(3,0)}$\\$+\lambda \phi_{(1,2)}^{\dagger}(\tilde{H}^{\dagger}\sigma_aH)\phi_{(3,0)}$\\$+y_1 \overline{\ell_L^c} \phi_{(1,2)} \ell_L$}& no &\parbox[c]{3cm}{discussed in~\cite{Law2013};\\no proper mixing;\\ruled out} \\ \hline
A5& $\phi_{(1,2)}$, $\phi_{(4,1)}$ &1 & 9 & \parbox[c]{3.7cm}{$\overline{\ell_L^c} \phi_{(1,2)}\ell_L$\\$+\lambda\phi_{(4,1)}^{\dagger}HH^\dagger H$\\$+\mu_1\tilde{\phi}_{(4,1)}^\dagger \phi_{(4,1)} \tilde{\phi}_{(1,2)}$} & no & discussed in~\cite{Law2013}\\ \hline
\caption{Radiative neutrino mass models with two scalars without colour.}
\label{tab:A}
\end{longtable}

The first model with two new representations (A1) is the Zee-Babu model~\cite{Zee1985, Babu1988}. It leads to a symmetric neutrino mass matrix with respect to the family indices with one copy of each of the new fields.

The model (A2) was discussed in Ref.~\cite{Law2013} as a simplification of the Zee-model. Compared to the Zee-model it is more restrictive, since there are less new couplings of the Higgs boson and the new $SU(2)_L$-doublet in the scalar potential. As a result, the mass matrix is traceless in flavour space and does not produce the correct mixing. This minimal model is therefore ruled out.

Similar to (A0), the model (A3) only works in the presence of two Higgs doublets. Otherwise the term $H^{\dagger}\sigma_a\phi_{(2,3)}H^\dagger \sigma_a\tilde{H}$ vanishes identically and no neutrino mass is generated. This model was also discussed in Ref.~\cite{Law2013}.

The models (A1)-(A3) are all two-loop realisations of the dimension-5 Weinberg operator. In contrast, the model (A4) is a one-loop realisation of the dimension-7 operator $(\ell_L H H \ell_L)(H^\dagger H)/\Lambda_L^3$. However, the mass matrix is antisymmetric in flavour space and no proper mixing is generated~\cite{Law2013}. This minimal model is therefore also ruled out.

The final model with two new colour-neutral scalars produces the dimension-9 operator $(\ell_L H H \ell_L)(H^\dagger H)(H^\dagger H)/\Lambda_L^5$ at one-loop level~\cite{Law2013}. This models requires two copies of the quadruplet scalar since the coupling $\tilde{\phi}_{(4,1)}^\dagger \phi_{(4,1)} \tilde{\phi}_{(1,2)}$ vanishes otherwise, and therefore also the neutrino mass. Moreover, a variation of the model where only one of the copies of $\phi_{(4,1)}$ acquires a vacuum expectation value can be excluded, since it produces only a traceless neutrino mass matrix.

\subsubsection{Minimal models with one new scalar and one new fermion without colour}
In Table~\ref{tab:BC}, we list all minimal models generating neutrino masses at loop-level with a new scalar and a new fermionic field transforming trivially with respect to $SU(3)_C$. In general, for models with one new fermion and one new scalar, one needs either two copies of the fermion or of the scalar to produce at least two independent non-zero neutrino masses.
\enlargethispage{0.5cm}
\begin{longtable}{|c|p{3cm}|c|c|c|c|p{3.5cm}|}
\hline
Model & New fields & Loops & Eff. dim. & Relevant interactions & New? & Comments \\ \hline \endhead
B1 & \parbox[c]{2.7cm}{$\phi_{(1,2)}$,\\ $\psi^{(2,-3)}_L+\psi^{(2,-3)}_R$} & 2 & 5 & \parbox[c]{3.5cm}{$y\overline{\ell_L^c}\phi_{(1,2)} \ell_L$ \\$+y\overline{\psi^{(2,-3)}_R}\ell_L \tilde{\phi}_{(1,2)}$\\$+y \overline{e_R}\psi^{(2,-3)}_LH$\\$+m\overline{\psi^{(2,-3)}_L}\psi^{(2,-3)}_R$} & no &  \parbox[c]{3.5cm}{introduced in~\cite{Cai2014}\\no proper mixing;\\ruled out}\\ \hline
B2 & \parbox[c]{2.7cm}{$\phi_{(2,3)}$,\\ $\psi^{(1,-2)}_L+\psi^{(1,-2)}_R$} & 1 & 5 & \parbox[c]{3.5cm}{$y\overline{\ell_L^c}\phi_{(2,3)} \psi^{(1,-2)}_L$ \\$+y\overline{\ell_L} H \psi^{(1,-2)}_R $\\$+\lambda H^{\dagger}\phi_{(2,3)}H^\dagger \tilde{H}$\\$+m\overline{\psi^{(1,-2)}_L}\psi^{(1,-2)}_R$} & yes & requires two Higgs doublets\\ \hline
B3 & \parbox[c]{2.7cm}{$\phi_{(2,3)}$,\\ $\psi^{(2,-1)}_L+\psi^{(2,-1)}_R$ }& 2 & 5 & \parbox[c]{3.5cm}{$y\overline{e_R^c}\phi_{(2,3)} \psi^{(2,-1)}_R$ \\$+y\overline{e_R} \tilde{H} \psi^{(2,-1)}_L $\\$+\lambda H^{\dagger}\phi_{(2,3)}H^\dagger \tilde{H}$\\$+m\overline{\psi^{(2,-1)}_L}\psi^{(2,-1)}_R$} & yes & requires two Higgs doublets\\ \hline
B4 & \parbox[c]{2.7cm}{$\phi_{(2,3)}$,\\ $\psi^{(3,-2)}_L+\psi^{(3,-2)}_R$} & 1& 5 & \parbox[c]{3.5cm}{$y\overline{\ell_L^c}\phi_{(2,3)} \psi^{(3,-2)}_L$ \\$+\lambda H^{\dagger}\tilde{H}H^{\dagger}\phi_{(2,3)}$\\$+\overline{\ell_L}H\psi^{(3,-2)}_R$\\$+m\overline{\psi^{(3,-2)}_L}\psi^{(3,-2)}_R$} & yes & requires two Higgs doublets\\ \hline
B5 & \parbox[c]{2.7cm}{$\phi_{(4,3)}$,\\ $\psi^{(3,-2)}_L+\psi^{(3,-2)}_R$} & 1& 5 & \parbox[c]{3.5cm}{$y\overline{\ell_L^c}\phi_{(4,3)} \psi^{(3,-2)}_L$ \\$+\lambda H^{\dagger}\tilde{H}H^{\dagger}\phi_{(4,3)}$\\$+\overline{\ell_L}H\psi^{(3,-2)}_R$\\$+m\overline{\psi^{(3,-2)}_L}\psi^{(3,-2)}_R$} & no & \parbox[c]{3.5cm}{ discussed in~\cite{Babu2009}\\with two copies of\\fermions}\\ \hline 
C1 &$\phi_{(4,1)}$, $\psi^{(5,0)}_R$ & 1 & 5 & \parbox[c]{3.5cm}{$M\overline{(\psi^{(5,0)}_R)^c}\psi^{(5,0)}_R$\\$+y\overline{\ell_L} \phi_{(4,1)}^{\dagger}\psi^{(5,0)}_R$\\$+\lambda H^{\dagger}HH^{\dagger}\phi_{(4,1)}$} & no & discussed by~\cite{Liao2010, Kumericki2012, Picek2012, McDonald2013, Chen2013, Ding2014} \\ \hline
C2 &\parbox[c]{2.7cm}{ $\phi_{(n,1)}$, $\psi^{(n\pm1,0)}_R$,\\  $n>4$,  $n$ even} & 1 & 5 &\parbox[c]{3.5cm}{ $M\overline{(\psi^{(n\pm1,0)}_R)^c}\psi^{(n\pm1,0)}_R$\\$+y\overline{\ell_L} \tilde{\phi}_{(n,1)}\psi^{(n\pm1,0)}_R$\\$+\lambda H^{\dagger}\phi_{n,1}H^{\dagger}\phi_{n,1}$} & no&\parbox[c]{3.5cm}{ case $\phi_{(6,1)}+\psi^{(5,0)}_R$\\ discussed in~\cite{Cai2011}; case\\ $\phi_{(3,1)}+\psi^{(2,0)}_R$ discussed \\in~\cite{Perez2009}}\\ \hline 
\caption{Radiative neutrino mass models with one scalar and one fermion without colour.}
\label{tab:BC}
\end{longtable}
The first model with one new scalar and a new vector-like fermion, (B1), was found in Ref.~\cite{Cai2014} in a study on dimension-7 effective operators. It is a tree-level realisation of the dimension-7 operator $\ell_L\ell_L \ell_L \overline{e_R} H/\Lambda_L^3$, closed off at two-loop to the dimension-5 Weinberg operator by connecting $\overline{e_R}$ and $\ell_L$. The new fermion enables the coupling $\mu \phi_{(1,2)}^\dagger \tilde{H}^\dagger H$ via a fermion loop. As a result, the neutrino mass matrix will be traceless, as in the case of (A2), and no proper mixing will be generated. This model is therefore ruled out.

The models (B2) and (B3) have not been discussed previously to the best of our knowledge. Both of them require two Higgs doublets. With only one Higgs the coupling $ H^{\dagger}\phi_{(2,3)}H^\dagger \tilde{H}$ vanishes identically, leading to a vanishing neutrino mass. While (B2) realises neutrino masses at one-loop level, (B3) generates them at two-loop level.

The model (B4) is the triplet-analogue to (B2). Hence, it also requires two copies of the Standard Model Higgs and induces radiative neutrino masses at one-loop. This model has not been discussed previously.

The model (B5) was introduced and studied in Ref.~\cite{Babu2009} in a version with two vector-like copies of the fermion field. The second copy of the fermion is necessary to produce the correct phenomenology. Depending on the masses of the fields, the dominant contribution can be either the dimension-5 operator  generated at one-loop level or the tree-level realisation of the dimension-7 operator $(\ell_L\ell_L HH)(H^\dagger H)/\Lambda_L^3$.

The model (C1) was introduced in Ref.~\cite{Liao2010} and studied in Ref.~\cite{Ding2014} as the starting point of a chain of models introducing higher dimensional operators at tree-level as their leading term. It realises the dimension-9 operator $\ell_L \ell_L HH(H^\dagger H)(H^\dagger H)/\Lambda_L^5$ at tree-level. The dimension-5 Weinberg operator appears at one-loop.

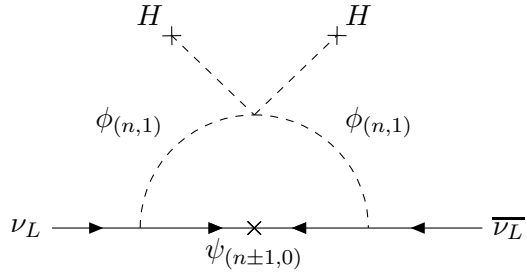
\begin{figure}[!ht]
\centering
\scalebox{1.0}{
\begin{tikzpicture}
\begin{feynman}
\vertex (a){\(\nu_L\)};
\vertex[right=of a](x);
\vertex[right=of x](y);
\vertex[right=of y](z);
\vertex[right=of z](b){\(\overline{\nu_L}\)};
\vertex[above=of y](w);
\vertex[above left=of w](c){\(H\)};
\vertex[above right=of w](d){\(H\)};
\diagram*[small, horizontal=a to b]{
(a)--[fermion](x)--[majorana, edge label'=\(\psi_{(n\pm 1,0)}\), insertion=0.5](z)--[anti fermion](b),
(x)--[scalar, quarter left, edge label=\(\phi_{(n,1)}\)](w)--[scalar, quarter left, edge label=\(\phi_{(n,1)}\)](z),
(c)--[scalar, insertion=0.01](w)--[scalar, insertion=0.99](d),
};
\end{feynman}
\end{tikzpicture}
}
\caption{Neutrino mass diagram resulting from model class C2.}
\label{fig:B5}
\end{figure}

(C2) is a class of models that all generate neutrino mass via the same mechanism. It contains the new scalar field $\phi_{(n,1)} \sim (1,n,1)$ and the Majorana fermion $\psi_{n\pm 1,0}= \psi^{(n\pm 1,0)}_R+(\psi^{(n\pm 1,0)}_R)^c \sim(1,n\pm 1,0)$, where $n>4$, and n even.  The relevant interactions are then given by 
\begin{equation}
\mathcal{L} \supset M\overline{\psi_{n\pm1,0}^c}\psi_{n\pm1,0}+y\overline{\ell}_L \tilde{\phi}_{n,1}\psi_{n\pm1,0}+\lambda H^{\dagger}\phi_{n,1}H^{\dagger}\phi_{n,1} \, .
\end{equation}
The resulting mass diagram is shown in Fig.~\ref{fig:B5}.

The case with $\phi_{(3,1)} \sim (1,3,1)$ and $\psi^{(2,0)}_R \sim (1,2,0)$ was discussed in Ref.~\cite{Perez2009}. 
However, it seems that in this case no neutrino mass is generated. The singlet from two fields in an even $SU(2)$ representation is always a completely antisymmetric combination of the fields. So at least two flavours of the $\psi^{(2,0)}_R$ are needed. Taking into account the anticommutiation of fermion fields, the Majorana mass matrix of these new particles is antisymmetric in flavour space. Since the rest of the interaction is totally symmetric, the resulting mass matrix for the neutrinos will be antisymmetric, leading to a vanishing Majorana mass.

Note that this model contains a potential dark matter candidate -- the neutral component of the Majorana fermion. The example with $\phi_{(6,1)} \sim (1,6,1)$ and $\psi^{(5,0)}_R \sim (1,5,0)$ was studied in Ref.~\cite{Cai2011}. It was discussed as a model generating neutrino masses radiatively and containing a viable dark matter candidate due to an accidental $Z_2$ symmetry.

\subsection{Minimal radiative neutrino mass models with colour}
\label{sec:colour}

In this section, we consider minimal models where one or both new representations transform non-trivially with respect to $SU(3)_C$. We define the representations of $SU(3)_C$ such that $3 \otimes 3 \sim \bar{6} \oplus \bar{3}$. Note that now the conjugation of the field does not only change the sign of the hypercharges, but also the $SU(3)_C$-representation from $R$ to $\bar{R}$, where $R$ is any complex representation of $SU(3)_C$.

In order to break lepton number with new coloured scalar fields, we have to look at fermion bilinears of Standard Model quarks and leptons and all possible quark couplings which can inverse the fermion number flow. For new coloured fermions, we take all couplings of Standard Model quarks and the Higgs field into account. We find several coloured models of type A, called type cA, as well as coloured B-type (cB) and C-type (cC) models which are given in the next two subsections. Note that, apart from lepton number violation, some of the models also introduce baryon number violation. To avoid fast proton decays in these cases, the new particles have to be sufficiently heavy.

\subsubsection{Minimal models with two new scalars with colour}

In Table~\ref{tab:cA}, we list the minimal models which can generate neutrino masses at loop-level with two new scalar fields where at least one scalar field transforms in a non-trivial representation of $SU(3)_C$. There are basically two categories of models, one (cA1, cA3, cA5, cA7) with hypercharges $4/3$ and $-2/3$ and one (cA2, cA4, cA6, cA8) with hypercharges $1/3$ and $-2/3$. Note that due to the symmetry properties of the representations used, the models cA1, cA3, cA4, and cA6 need two copies of the new scalar with lower hypercharge to achieve non-vanishing neutrino masses. Otherwise the lepton number violating term of the form $\phi_a \phi_b^2$, where $\phi_b$ represents the new scalar with lower hypercharge, vanishes.

\begin{longtable}{|c|c|c|c|c|p{4cm}|}
\hline
Model & New Fields & Interactions & B-viol?&New? & Comment\\ \hline \endhead
cA1 & $\phi_{(3,1,-2/3)}$, $\phi_{(3,1,4/3)}$ & \parbox[c]{4.95cm}{$y_1 \overline{\ell_L^c} Q_L \phi_{(3,1,-2/3)}^\dagger $\\ $+y_2 \overline{Q^c_L} \phi_{(3,1,-2/3)} Q_L$\\$+y_3 \overline{d^c_R} \phi_{(3,1,4/3)} d_R$\\$+\mu\phi_{(3,1,-2/3)} \phi_{(3,1,-2/3)} \phi_{(3,1,4/3)} $}& yes & yes &\parbox[c]{4cm}{scalar up- and down-\\quark; two copies of\\$\phi_{(3,1,-2/3)}$ required}\\ \hline
cA2 &  $\phi_{(3,1,-2/3)}$, $\phi_{(3,2,1/3)}$ & \parbox[c]{4.95cm}{$y_1\overline{\ell_L}d_R\phi_{(3,2,1/3)}^\dagger$\\ $+y_2\overline{Q^c_L} \phi_{(3,1,-2/3)}Q_L$\\$+\mu\phi_{(3,1,-2/3)}\phi_{(3,2,1/3)}\phi_{(3,2,1/3)}$}& yes& no & \parbox[c]{4cm}{discussed in~\cite{Babu2001, Babu2010};\\ embedding in $SU(5)$ \\and $SO(10)$~\cite{Dorsner2017}}\\ \hline
cA3 &  $\phi_{(3,3,-2/3)}$, $\phi_{(3,1,4/3)}$ & \parbox[c]{4.95cm}{$y_1 \overline{\ell^c_L} Q_L \phi_{(3,3,-2/3)}^\dagger $\\ $+y_2 \overline{Q^c_L} \phi_{(3,3,-2/3)} Q_L$\\$+y_3 \overline{d^c_R} \phi_{(3,1,4/3)} d_R$\\$+\mu\phi_{(3,3,-2/3)}\phi_{(3,3,-2/3)}\phi_{(3,1,4/3)}$}& yes &yes&\parbox[c]{4cm}{scalar up-quark and\\triplet version of scalar\\down-quark; two copies of\\$\phi_{(3,3,-2/3)}$ required}\\ \hline
cA4 &  $\phi_{(3,3,-2/3)}$, $\phi_{(3,2,1/3)}$ & \parbox[c]{4.95cm}{$y_1\overline{\ell_L}d_R\phi_{(3,2,1/3)}^\dagger$\\ $+y_2\overline{Q^c_L}\phi_{(3,3,-2/3)}Q_L$\\$+\mu\phi_{(3,3,-2/3)}\phi_{(3,2,1/3)}\phi_{(3,2,1/3)}$}&yes&no & \parbox[c]{4cm}{mentioned in~\cite{Cai2014};\\embedding in $SU(5)$ and\\$SO(10)$~\cite{Dorsner2017}; discussed in\\the context of the $R_K$\\anomaly~\cite{Pas2015, Cheung2016};\\two copies of $\phi_{(3,2,1/3)}$\\required}\\ \hline
cA5 &  $\phi_{(3,1,-2/3)}$, $\phi_{(6,1,4/3)}$ & \parbox[c]{4.95cm}{$y_1 \overline{\ell_L^c} Q_L \phi_{(3,1,-2/3)}^\dagger $\\ $+y_2 \overline{Q^c_L} \phi_{(3,1,-2/3)} Q_L$\\$+y_3 \overline{d^c_R} \phi_{(6,1,4/3)}d_R$\\$+\mu\phi_{(3,1,-2/3)}\phi_{(3,1,-2/3)}\phi_{(6,1,4/3)}$}& yes& no & discussed in~\cite{Chang2016}\\ \hline
cA6 &  $\phi_{(6,1,-2/3)}$, $\phi_{(3,2,1/3)}$ & \parbox[c]{4.95cm}{$y_1\overline{\ell_L}d_R\phi_{(3,2,1/3)}^\dagger$\\ $+y_2\overline{Q^c_L}\phi_{(6,1,-2/3)}Q_L$\\$+\mu\phi_{(6,1,-2/3)}\phi_{(3,2,1/3)}\phi_{(3,2,1/3)}$}& no& yes &\parbox[c]{4cm}{scalar quark doublet and\\colour sextet; two copies \\of $\phi_{(3,2,1/3)}$ required}\\ \hline
cA7 &  $\phi_{(3,3,-2/3)}$, $\phi_{(6,1,4/3)}$ & \parbox[c]{4.95cm}{$y_1\overline{\ell_L^c}Q_L\phi_{(3,3,-2/3)}^\dagger$\\ $+y_2\overline{Q^c_L} \phi_{(3,3,-2/3)} Q_L$\\$+y_3\overline{d^c_R} \phi_{(6,1,4/3)} d_R$\\$+\mu\phi_{(3,3,-2/3)}\phi_{(3,3,-2/3)}\phi_{(6,1,4/3)}$}& yes&yes & variation of cA1\\ \hline
cA8 &  $\phi_{(6,3,-2/3)}$, $\phi_{(3,2,1/3)}$ & \parbox[c]{4.95cm}{$y_1\overline{\ell_L}d_R\phi_{(3,2,1/3)}^\dagger$\\ $+y_2\overline{Q_L^c} \phi_{(6,3,-2/3)}Q_L$\\$+\mu\phi_{(6,3,-2/3)}\phi_{(3,2,1/3)}\phi_{(3,2,1/3)}$}& no& yes &\parbox[c]{4cm}{ scalar quark doublet and\\exotic colour sextet with\\non-trivial $SU(2)_L$ charge}\\ \hline 
\caption{List of neutrino mass models with two new scalars including non-trivial $SU(3)_C$-representations.}
\label{tab:cA}
\end{longtable}
It is interesting to note that scalar leptoquarks have also recently been considered in the context of the measured B-anomalies. The hints of lepton universality violations by measurements of $R_K = \mathcal{B}(\bar{B}\to \bar{K}\mu \mu)/\mathcal{B}(\bar{B}\to \bar{K} e e)$~\cite{Aaij2014} can be explained by introducing the scalar leptoquark $\phi_{(3,2,1/3)}$~\cite{Hiller2014, Becirevic2016, Becirevic2016b} where Ref.~\cite{Hiller2014} also considers $\phi_{(3,3, -2/3)}$. Additionally, the scalar leptoquark $\phi_{(3,1, -2/3)}$ is also considered to explain the measured $R_{D^{(*)}} = \mathcal{B}(\bar{B}\to D^{(*)}\tau \bar{\nu})/\mathcal{B}(\bar{B}\to D^{(*)}\ell \bar{\nu})$~\cite{Lees2013, Huschle2015, Aaij2015, Fajfer2012, Becirevic2012} anomaly by Ref.~\cite{Bauer2015, Becirevic2016b}. Note however that the scalar leptoquark $\phi_{(3,1, -2/3)}$ leads to proton decay and therefore $M_{\phi_{(3,1, -2/3)}} \gtrsim 3 \cdot 10^{11}$ GeV~\cite{Nath2006}\footnote{This limit contains the assumption that the scalar leptoquark couplings are of order $10^{-5}$. For couplings of order one, the bound can be as high as \SI{e15}{GeV}.}. This bound should apply similarly to models containing the scalar leptoquark $\phi_{(3,3, -2/3)}$. A thorough phenomenological study of possible connections of the B-anomalies and neutrino masses is, however, beyond the scope of this work.

\subsubsection{Minimal models with one new scalar and one new fermion with colour}
\label{sec:c1S1F}
Table~\ref{tab:cB} lists all models of type cB where neutrino masses are generated by a new scalar representation and a new Dirac fermion with at least one new coloured field.
\begin{longtable}{|c|c|c|c|c|p{4.5cm}|}
\hline
Model & New Fields & Interactions & B-viol?&New? & Comment\\ \hline \endhead
cB1 & \parbox[c]{3.7cm}{$\phi_{(1,1,2)}$,\\ $\psi^{(3,1,-2/3)}_L+\psi^{(3,1,-2/3)}_R$} & \parbox[c]{4.5cm}{$y_1 \overline{\ell^c_L}\phi_{(1,1,2)}\ell_L$\\$+y_2\overline{Q_L} H \psi^{(3,1,-2/3)}_R$\\$+y_3\overline{u_R}\phi_{(1,1,2)}\psi^{(3,1,-2/3)}_L$\\$+m\overline{\psi^{(3,1,-2/3)}_L}\psi^{(3,1,-2/3)}_R$}&no&no &\parbox[c]{4.5cm}{ mentioned in~\cite{Cai2014}; singly\\charged scalar and vector-like\\down-type quark\\no proper mixing;\\ruled out}\\ \hline
cB2 & \parbox[c]{3.7cm}{$\phi_{(1,1,2)}$,\\ $\psi^{(3,1,4/3)}_L+\psi^{(3,1,4/3)}_R$ }& \parbox[c]{4.5cm}{$y_1 \overline{\ell^c_L}\phi_{(1,1,2)}\ell_L$\\$+y_2\overline{Q_L} H^\dagger \psi^{(3,1, 4/3)}_R$\\$+y_3\overline{d_R}\phi_{(1,1,2)}^\dagger \psi^{(3,1,4/3)}_L$\\$+m\overline{\psi^{(3,1,4/3)}_L}\psi^{(3,1,4/3)}_R$}&no&no &\parbox{4.5cm}{ mentioned in~\cite{Cai2014}; singly\\charged scalar and vector-like\\up-type quark\\no proper mixing;\\ruled out} \\ \hline
cB3 & \parbox[c]{3.7cm}{$\phi_{(1,1,2)}$,\\ $\psi^{(3,2,-5/3)}_L+\psi^{(3,2,-5/3)}_R$} & \parbox[c]{4.5cm}{$y_1 \overline{\ell^c_L}\phi_{(1,1,2)}\ell_L$\\$+y_2\overline{d_R} H \psi^{(3,2, -5/3)}_L$\\$+y_3\overline{Q_L}\phi_{(1,1,2)} \psi^{(3,2,-5/3)}_R$\\$+m\overline{\psi^{(3,2,-5/3)}_L}\psi^{(3,2,-5/3)}_R$}&no&no &\parbox[c]{4.5cm}{mentioned in~\cite{Cai2014}\\no proper mixing;\\ruled out}\\ \hline
cB4 & \parbox[c]{3.7cm}{$\phi_{(1,1,2)}$,\\ $\psi^{(3,2,7/3)}_L+\psi^{(3,2,7/3)}_R$} & \parbox[c]{4.5cm}{$y_1 \overline{\ell^c_L}\phi_{(1,1,2)}\ell_L$\\$+y_2\overline{u_R} H^\dagger \psi^{(3,2, 7/3)}_L$\\$+y_3\overline{Q_L}\phi_{(1,1,2)}^\dagger \psi^{(3,2,7/3)}_R$\\$+m\overline{\psi^{(3,2,7/3)}_L}\psi^{(3,2,7/3)}_R$}&no&no & \parbox[c]{4.5cm}{considered in~\cite{Cai2014}\\no proper mixing;\\ruled out}\\ \hline
cB5 & \parbox[c]{3.7cm}{$\phi_{(3,2,1/3)}$,\\ $\psi^{(1,2,1)}_L+\psi^{(1,2,1)}_R$} & \parbox[c]{4.5cm}{$y_1 \overline{\ell_L} \phi_{(3,2,1/3)}^\dagger d_R$\\$+y_2 \overline{u_R} \phi_{(3,2,1/3)} \psi^{(1,2,1)}_L$\\$+y_3 \overline{e_R^c} H \psi^{(1,2,1)}_R$\\$+m\overline{\psi^{(1,2,1)}_L}\psi^{(1,2,1)}_R$} & yes & no & \parbox[c]{4.5cm}{mentioned in~\cite{Cai2014}; scalar\\quark doublet and vector-like\\lepton doublet; gauge\\unification at $10^{14}$ GeV~\cite{Hagedorn2016}} \\ \hline
cB6 & \parbox[c]{3.7cm}{$\phi_{(3,2,1/3)}$,\\ $\psi^{(3,1,-2/3)}_L+\psi^{(3,1,-2/3)}_R$} & \parbox[c]{4.5cm}{$y_1 \overline{\ell_L} \phi_{(3,2,1/3)}^\dagger d_R$\\$+y_2\overline{\ell_L} \phi_{(3,2,1/3)}^\dagger \psi^{(3,1,-2/3)}_R$\\$+y_3\overline{Q_L^c}\phi_{(3,2,1/3)} \psi^{(3,1,-2/3)}_L$\\$+\lambda H^\dagger \phi_{(3,2,1/3)}^3$\\$+m\overline{\psi^{(3,1,-2/3)}_L}\psi^{(3,1,-2/3)}_R$} & yes & yes & \parbox[c]{4.5cm}{scalar quark doublet and\\vector-like down-type quark;\\ two copies of $\phi_{(3,2,1/3)}$\\required} \\ \hline
cB7 & \parbox[c]{3.7cm}{$\phi_{(3,2,1/3)}$,\\ $\psi^{(3,1,4/3)}_L+\psi^{(3,1,4/3)}_R$} & \parbox[c]{4.5cm}{$y_1\overline{\ell_L} \phi_{(3,2,1/3)}^\dagger d_R$\\$+y_2\overline{\ell_L^c}\phi_{(3,2,1/3)}^\dagger \psi^{(3,1,4/3)}_L$\\$+y_3\overline{Q_L} H^\dagger \psi^{(3,1,4/3)}_R$\\$+m\overline{\psi^{(3,1,-2/3)}_L}\psi^{(3,1,-2/3)}_R$} & yes & no &\parbox[c]{4.5cm}{ discussed in~\cite{Babu2011}; scalar quark\\doublet and vector-like\\up-type quark} \\ \hline
cB8 & \parbox[c]{3.7cm}{$\phi_{(3,1,-2/3)}$, \\$\psi^{(3,2,-5/3)}_L+\psi^{(3,2,-5/3)}_R$} & \parbox[c]{4.5cm}{$y_1\overline{\ell^c_L} \phi_{(3,1,-2/3)}^\dagger Q_L$\\$+y_2\overline{\ell_L}\phi_{(3,1,-2/3)}^\dagger \psi^{(3,2,-5/3)}_R$\\$+y_3\overline{d_R} H \psi^{(3,2,-5/3)}_L$\\$+m\overline{\psi^{(3,2,-5/3)}_L}\psi^{(3,2,-5/3)}_R$} & yes & no & discussed in~\cite{Cai2014, Popov2016} \\ \hline
cB9 & \parbox[c]{3.7cm}{$\phi_{(3,2,1/3)}$,\\ $\psi^{(3,2,1/3)}_L+\psi^{(3,2,1/3)}_R$} & \parbox[c]{4.5cm}{$y_1\overline{\ell_L} \phi_{(3,2,1/3)}^\dagger d_R$\\$+y_2\overline{d_R^c}\phi_{(3,2,1/3)}\psi^{(3,2,1/3)}_R$\\$+y_3\overline{d_R} H^\dagger \psi^{(3,2,1/3)}_L$\\$+\lambda H^\dagger \phi_{(3,2,1/3)}^3$\\$+m\overline{\psi^{(3,2,1/3)}_L}\psi^{(3,2,1/3)}_R$} & yes & yes & \parbox[c]{4.5cm}{scalar quark doublet and\\vector-like quark doublet;\\two copies of $\phi_{(3,2,1/3)}$\\required}\\ \hline
cB10 & \parbox[c]{3.7cm}{$\phi_{(3,2,1/3)}$,\\ $\psi^{(3,2,7/3)}_L+\psi^{(3,2,7/3)}_R$} & \parbox[c]{4.5cm}{$y_1\overline{\ell_L}\phi_{(3,2,1/3)}^\dagger d_R$\\$+y_2\overline{e_R^c}\phi_{(3,2,1/3)}^\dagger \psi^{(3,2,7/3)}_R$\\$+y_3\overline{u_R} H^\dagger \psi^{(3,2,7/3)}_L$\\$+m\overline{\psi^{(3,2,7/3)}_L}\psi^{(3,2,7/3)}_R$} & yes & no & mentioned in~\cite{Cai2014} \\ \hline
cB11 & \parbox[c]{3.7cm}{$\phi_{(3,2,1/3)}$,\\ $\psi^{(3,3,-2/3)}_L+\psi^{(3,3,-2/3)}_R$} & \parbox[c]{4.5cm}{$y_1 \overline{\ell_L} \phi_{(3,2,1/3)}^\dagger d_R$\\$+y_2\overline{\ell_L} \phi_{(3,2,1/3)}^\dagger \psi^{(3,3,-2/3)}_R$\\$+y_3\overline{Q_L^c}\phi_{(3,2,1/3)} \psi^{(3,3,-2/3)}_L$\\$+\lambda H^\dagger \phi_{(3,2,1/3)}^3$\\$+m\overline{\psi^{(3,3,-2/3)}_L}\psi^{(3,3,-2/3)}_R$} & yes & yes & \parbox[c]{4.5cm}{scalar quark doublet and\\triplet version of down-quark;\\two copies of $\phi_{(3,2,1/3)}$\\ required}\\ \hline
cB12 & \parbox[c]{3.7cm}{$\phi_{(3,2,1/3)}$, \\$\psi^{(3,3,4/3)}_L+\psi^{(3,3,4/3)}_R$} & \parbox[c]{4.5cm}{$y_1\overline{\ell_L}\phi_{(3,2,1/3)}^\dagger d_R$\\$+y_2\overline{\ell_L^c}\phi_{(3,2,1/3)}^\dagger \psi^{(3,3,4/3)}_L$\\$+y_3\overline{Q_L} H^\dagger \psi^{(3,3,4/3)}_R$\\$+m\overline{\psi^{(3,3,4/3)}_L}\psi^{(3,3,4/3)}_R$} & yes & no & mentioned in~\cite{Cai2014} \\ \hline
cB13 & \parbox[c]{3.7cm}{$\phi_{(3,3,-2/3)}$,\\ $\psi^{(3,2,-5/3)}_L+\psi^{(3,2,-5/3)}_R$} & \parbox[c]{4.5cm}{$y_1\overline{\ell^c_L} \phi_{(3,3,-2/3)}^\dagger Q_L$\\$+y_2\overline{\ell_L} \phi_{(3,3,-2/3)}^\dagger \psi^{(3,2,-5/3)}_R$\\$+y_3\overline{d_R} H \psi^{(3,2,-5/3)}_L$\\$+m\overline{\psi^{(3,2,-5/3)}_L}\psi^{(3,2,-5/3)}_R$} & yes & no & listed in~\cite{Cai2014} \\ \hline
\pagebreak
cB14 & \parbox[c]{3.7cm}{$\phi_{(3,2,1/3)}$,\\ $\psi^{(8,2,1)}_L+\psi^{(8,2,1)}_R$} & \parbox[c]{4.5cm}{$y_1 \overline{\ell_L} \phi_{(3,2,1/3)}^\dagger d_R$\\$+y_2 \overline{u_R} \phi_{(3,2,1/3)} \psi^{(8,2,1)}_L$\\$+y_3 \overline{d_R^c} \phi_{(3,2,1/3)}^\dagger  \psi^{(8,2,1)}_R$\\$+m\overline{\psi^{(8,2,1)}_L}\psi^{(8,2,1)}_R$} & yes & yes & colour octett version of cB5 \\ \hline 
\caption{Neutrino mass models with one new scalar and one new vector-like Dirac fermion with at least one coloured field.}
\label{tab:cB}
\end{longtable}
The models (cB1)-(cB4) are variants of the model (B1) with coloured fields running in the fermion loop. As a result, they are also ruled out due to the resulting traceless neutrino mass matrix. 

Models of type cC where a new scalar field and a Majorana fermion with at least one coloured field generate neutrino masses at loop-level are included in Table~\ref{tab:cC}. These encompass the coloured seesaws~\cite{Perez2009}, as well as models containing gluinos and squarks.

\begin{longtable}{|c|c|c|c|c|p{4cm}|}
\hline
Model & New Fields & Interactions & B-viol?&New? & Comment\\ \hline \endhead
cC1 & $\phi_{(3,1,-2/3)}$, $\psi^{(8,1,0)}_L$ & \parbox[c]{4.3cm}{$y_1 \overline{\ell^c_L} \phi_{(3,1,-2/3)}^\dagger Q_L$\\$+y_2 \overline{d_R} \phi_{(3,1,-2/3)} \psi^{(8,1,0)}_L$\\$+M \overline{(\psi^{(8,1,0)}_L)^c} \psi^{(8,1,0)}_L$} & yes & no & \parbox[c]{4cm}{discussed in~\cite{Angel2013, Cai2017a}; gluino \\and down-type squark} \\ \hline
cC2 & $\phi_{(3,2,1/3)}$, $\psi^{(8,1,0)}_R$ & \parbox[c]{4.3cm}{$y_1 \overline{\ell_L} \phi_{(3,2,1/3)}^\dagger d_R$\\$+y_2 \overline{Q_L} \phi_{(3,2,1/3)} \psi^{(8,1,0)}_R$\\$+M \overline{(\psi^{(8,1,0)}_R)^c} \psi^{(8,1,0)}_R$} & no & yes &  squark doublet and gluino \\ \hline
cC3 & $\phi_{(3,3,-2/3)}$, $\psi^{(8,3,0)}_L$ & \parbox[c]{4.3cm}{$y_1 \overline{\ell^c_L} \phi_{(3,3,-2/3)}^\dagger Q_L$\\$+y_2 \overline{d_R} \phi_{(3,3,-2/3)} \psi^{(8,3,0)}_L$\\$+M \overline{(\psi^{(8,3,0)}_L)^c} \psi^{(8,3,0)}_L$} & yes & yes & triplet version of cC1 \\ \hline
cC4& \parbox[c]{3.5cm}{$\phi_{(\mathcal{C},2,1)}$, $\psi^{(\mathcal{C},1,0)}_R$,\\ $\mathcal{C}$ real SU(3) rep.} & \parbox[c]{4.3cm}{$y_1 \overline{\ell_L} \phi_{(\mathcal{C},2,1)}^\dagger \psi^{(\mathcal{C},1,0)}_R$\\$+M \overline{(\psi^{(\mathcal{C},1,0)}_R)^c} \psi^{(\mathcal{C},1,0)}_R$\\$+\lambda H^\dagger \phi_{(\mathcal{C},2,1)} H^\dagger \phi_{(\mathcal{C},2,1)}$} & no & no & discussed in~\cite{Perez2009} \\ \hline
cC5 & \parbox[c]{3.5cm}{$\phi_{(\mathcal{C},2,1)}$, $\psi^{(\mathcal{C},3,0)}_R$,\\ $\mathcal{C}$ real SU(3) rep.} & \parbox[c]{4.3cm}{$y_1 \overline{\ell_L} \phi_{(\mathcal{C},2,1)}^\dagger \psi^{(\mathcal{C},3,0)}_R$\\$+M \overline{(\psi^{(\mathcal{C},3,0)}_R)^c} \psi^{(\mathcal{C},3,0)}_R$\\$+\lambda H^\dagger \phi_{(\mathcal{C},2,1)} H^\dagger \phi_{(\mathcal{C},2,1)}$} & no & no & discussed in~\cite{Perez2009} \\ \hline 
cC6 & \parbox[c]{3.5cm}{$\phi_{(\mathcal{C},n,1)}$, $\psi^{(\mathcal{C},n\pm 1,0)}_R$,\\ $\mathcal{C}$ real SU(3) rep.,\\ $n\geq 4$, $n$ even}&\parbox[c]{4.3cm}{$M\overline{(\psi^{(\mathcal{C},n\pm1,0)}_R)^c}\psi^{(\mathcal{C},n\pm1,0)}_R$\\$+y\overline{\ell_L} \phi_{(\mathcal{C},n,1)}^\dagger \psi^{(\mathcal{C},n\pm1,0)}_R$\\$+\lambda H^{\dagger}\phi_{(\mathcal{C},n,1)}H^{\dagger}\phi_{(\mathcal{C},n,1)}$}&no &yes & \parbox[c]{3.7cm}{ model C2 with coloured\\ fields; $n$ odd leads to\\fractional charges}\\ \hline
\caption{Neutrino mass models with one new scalar and one new Majorana fermion with at least one new coloured field.}
\label{tab:cC}
\end{longtable}
The models (cC1)~\cite{Angel2013} and (cC2) contain  particles which can be identified as gluinos and squarks. Hence, such models can be interesting to discuss in the context of supersymmetry. However, a realistic ultraviolet complete model is beyond the scope of this paper. Moreover, (cC3) is the triplet version of (cC1).

The models (cC4) and (cC5) are the coloured seesaw mechanisms. They use the fields $\phi_{(\mathcal{C},2,1)}$ and  $\psi^{(\mathcal{C},1,0)}_R$, or $\phi_{(\mathcal{C},2,1)}$ and $\psi^{(\mathcal{C},3,0)}_R$ where $\mathcal{C}$ denotes any real representation of $SU(3)_C$. Due to the colour charge, neutrino mass is generated at loop-level. They were discussed in Ref.~\cite{Perez2009}.

The model (cC6) is the generalisation of the model class (C2) to coloured models.
In contrast to (C2), this class of models does not contain fundamental dark matter candidates since the new fields carry colour charge. For recent work on coloured dark matter see Ref.~\cite{Gross2018, DeLuca2018}.

\section{Minimal radiative neutrino mass model with only new fermions}
\label{sec:minFerm}

In this section, we will first proof that there does not exist a Standard Model extension with two new fermionic representations which generate the dimension-5 Weinberg operator only at the quantum level without employing new symmetries. We will then present the unique model with three new fermions which generates the Weinberg operator at the two-loop level.

Let us assume there exist two fermions $\psi_a$ and $\psi_b$ that fulfil the requirements for inducing LNV. In other words, there exist three interactions involving those fermions that do not allow for a well-defined non-trivial lepton number assignment for $\psi_a$ and $\psi_b$. We can then differentiate three cases. 
\begin{enumerate}
\item Both fermions $\psi_a$ and $\psi_b$ have a direct coupling to Standard Model leptons. This strongly limits the possible representations of $\psi_a$ and $\psi_b$ to one of the following representations in the list
\begin{align}
\label{eq:SMcoupl}
\ell_L H &\sim (1,0)\oplus (3,0)\,, & e_R H &\sim (2,-1)\,, \nonumber \\
\ell^c_L H &\sim (1,2) \oplus (3,2)\,, & e_R^c H &\sim (2, 3)  \,,
\end{align}
or the corresponding field with opposite hypercharge.
Considering also the couplings of those fields to each other, one finds that no combination of any two of those fields leads to a viable model.
\item None of the fields couples directly to the Standard Model fermions. Since there are no bosons carrying lepton number in the Standard Model, $L(H)=L(W)=L(B)=L(G)=0$, there can at most be two equations for the lepton numbers of $\psi_a$ and $\psi_b$,
\begin{equation}
L(\psi_a)+L(\psi_b)=L(\psi_a)-L(\psi_b)=0 \, .
\end{equation} 
This can always be solved by setting $L(\psi_a)=L(\psi_b)=0$. There is no LNV in this case.
\item One of the fields, take $\psi_a$ w.l.o.g., couples directly to the Standard Model leptons, the other one, $\psi_b$, does not. Since we excluded $\nu_R \sim (1, 1, 0)$ and $\rho_R \sim (1, 3, 0)$ from these considerations as they lead to tree-level neutrino masses, it follows from the list~(\ref{eq:SMcoupl}) that $\psi_a$ has non-vanishing hypercharge. As a result either $\psi_a$ or $\psi_a^c$ can couple to the Standard Model leptons, but not both. This fixes the lepton number of $\psi_a$ to $L(\psi_a)=\pm1$. 

Next, we list all possible representations of $\psi_b$, that allow for a coupling to $\psi_a$, but not to the Standard Model. Among all representations coupling to the candidates for $\psi_a$ from~(\ref{eq:SMcoupl}) via a Yukawa coupling,
\begin{align}
\label{eq:candis}
\psi_{1,-2}H &\sim (2,-1)\,, & \psi_{1,-2}\tilde{H} &\sim (2,-3)\,,\\ \nonumber
\psi_{2,-1}H &\sim (1,0)\oplus (3,0)\,, & \psi_{2,-1}\tilde{H} &\sim (1,-2) \oplus (3,-2)\,, \\ \nonumber
\psi_{2,-3}H &\sim (1,-2) \oplus (3,-2)\,, & \psi_{2,-3}\tilde{H} &\sim (1,-4) \oplus (3,-4)\,,\\ \nonumber
\psi_{3,-2}H &\sim (2,-1) \oplus (4,-1)\,, & \psi_{3,-2} \tilde{H} &\sim (2,-3) \oplus (4,-3)\, , 
\end{align}
there are four such representations,
\begin{align}
\label{eq:coupl2}
\psi_B &\sim (1, 1, 4)\,, &\psi_B &\sim (1, 3, 4)\,, & \psi_B &\sim (1, 4, 1)\,, & \psi_B &\sim (1, 4, 3)\nonumber\, .
\end{align}
Now we have to find two interaction terms for $\psi_b$ that lead to a well-defined hypercharge compatible with the list of candidates, but an ill-defined lepton-number. The first possibility would be to consider the Majorana mass term $\overline{\psi_b^c}\psi_b$. It requires  $Y(\psi_b)=0$. However, there is no candidate with  $Y(\psi_b)=0$, so we exclude this case. The remaining possibility is having both,  the coupling of $\psi_b$ and $\psi_b^c$ to $\psi_a$. From them, we find the following conditions for the hypercharge of $\psi_a$ and $\psi_b$
\begin{align}
 Y(\psi_a)+Y(\psi_b)&= \pm 1 \,,\\
 Y(\psi_a)-Y(\psi_b)&= \pm 1\,,
\end{align} 
because the Higgs boson, the only way to mix fermions of different isospin or hypercharge in the Standard Model, has a hypercharge of one. Since the Higgs boson is colour neutral there is no way to mix fermions with different $SU(3)_C$-representations without adding a new scalar. This set of equations must be solved with $Y(\psi_a) = \pm1$. The only possible solution is again $Y(\psi_b)=0$. Therefore, there is no model with two new fermions. This concludes the proof.
\end{enumerate}

\begin{figure}[!ht]
\centering
\begin{tikzpicture}
\begin{feynman}
\vertex (a){\(\nu_L\)};
\vertex[right=of a](x);
\vertex[right=of x](i);
\vertex[right=of i](j);
\vertex[right=of j](m);
\vertex[right=of m](p);
\vertex[right=of p](y);
\vertex[right=of y](b){\(\overline{\nu_L}\)};
\vertex[below=of i](d){\(H\)};
\vertex[above=of p](e){\(H\)};
\diagram*[small, horizontal=a to b, inline=(a)]{
(a)--[fermion](x)--[fermion, edge label'=\(\psi_{3,-2}\)](i)--[fermion, edge label'=\(\psi_{4,-1}\)](j)--[majorana, edge label'=\(\psi_{5,0}\), insertion=0.5](m)--[anti fermion, edge label'=\(\overline{\psi}_{4,-1}\)](p)--[anti fermion, edge label'=\(\overline{\psi}_{3,-2}\)](y)--[anti fermion](b),
(x)--[scalar, half left, edge label=\(H\)](m),
(y)--[scalar, half left, edge label=\(H\)](j),
(i)--[scalar,  insertion=0.99](d),
(p)--[scalar,  insertion=0.99](e),};
\end{feynman}
\end{tikzpicture}
\caption{One possible realisation of the Weinberg operator in the unique model with only three new fermionic representations generating neutrino mass at loop level.}
\label{fig:C1}
\end{figure}
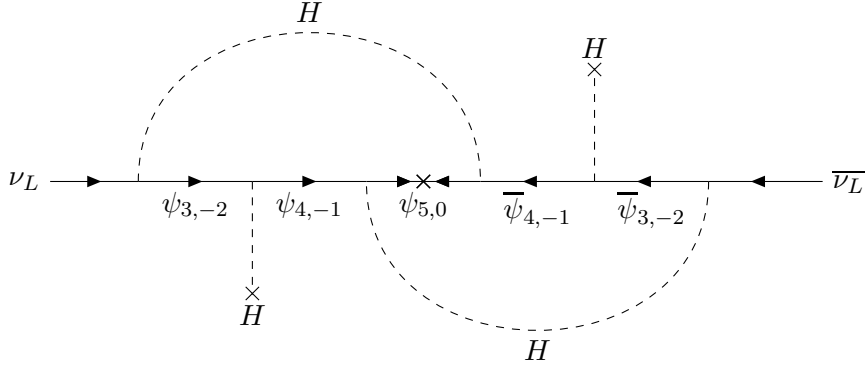

However, by going to three new fermionic representations, we find a unique model. Some aspects of this model were discussed in Ref.~\cite{Anamiati2018}. A proof of its uniqueness is given in the Appendix~\ref{app:proof}. The field content is given by two vector-like Dirac fermions
\begin{align}
\psi^{(3,-2)}_L+\psi^{(3,-2)}_R \sim (1,3,-2)\quad \text{and} \quad \psi^{(4,-1)}_L+\psi^{(4,-1)}_R \sim (1,4,-1)\,,
\end{align}
and a single Majorana fermion
\begin{align}
(\psi^{(5,0)}_R)^c+\psi^{(5,0)}_R\sim (1,5,0) \,.
\end{align}
We thus find the relevant interactions
\begin{align}
\mathcal{L} \supset M\overline{(\psi^{(5,0)}_R)^c}\psi^{(5,0)}_R+y_1\overline{\psi^{(3,-2)}}_R \tilde{H}\ell_L+y_2\overline{\psi^{(4,-1)}}_L H\psi^{(3,-2)}_R+y_3\overline{\psi^{(5,0)}}_R H\psi^{(4,-1)}_L \,.
\end{align}
Note that the dimension-5 operator is generated at two-loop level as can be inferred from Fig.~\ref{fig:C1}.

In the limit $y_1 \to 0$, this model has an accidental $Z_2$ symmetry
\begin{align}
\psi^{(3,-2)} &\to -\psi^{(3,-2)} \quad \text{and} \quad \overline{\psi^{(3,-2)}} \to -\overline{\psi^{(3,-2)}}\,, \nonumber \\
\psi^{(4,-1)} &\to -\psi^{(4,-1)} \quad \text{and} \quad \overline{\psi^{(4,-1)}} \to -\overline{\psi^{(4,-1)}}\,, \nonumber \\
\psi^{(5,0)} &\to -\psi^{(5,0)}\,.
\end{align}
It is thus technically natural that the Yukawa coupling $y_1$ is small. However, the smallness of this coupling does not only suppress the neutrino masses but also makes the lightest of the new fermionic particles approximately stable. If the lightest new fermion is electrically neutral, the presented model hence has a potential dark matter candidate. A thorough study of the connection of neutrino masses and dark matter in this scenario is left for future work.

Moreover, this model can also be viewed as a two-loop prototype version of a more general mechanism for neutrino mass suppression. One can write down such a mass model starting from any fermion in a $SU(2)_L\otimes U(1)_Y$-representation of the form (N,0), where $N$ is odd. To connect the Majorana fermion to the Standard Model leptons, an increasing number of intermediate new fermions is required. The higher the dimension $N$ of the representation of the Majorana fermion, the larger the loop order $l$ of the resulting mass diagram,
\begin{equation}
l=(N-3)\, , \, \, \, \, N\geq 3 .
\end{equation}
For large $N$ this will lead to a ladder structure in the mass diagram -- the representation of the fermion will increase up to $N$ and then decrease down to two again. Only neighbouring fermions in this diagram can directly couple to each other within the Standard Model. In this sense, this type of neutrino mass model resembles the Froggatt-Nielsen mechanism~\cite{Froggatt1978} and the Clockwork mechanism~\cite{Choi2015, Kaplan2015} -- there is a chain of particles, each of which can only interact with the neighbouring elements of the chain. This chain suppresses the introduced mass of the neutrino with respect to the mass of the Majorana fermion by including not only a Yukawa coupling, but also a loop order for each chain element.

\begin{figure}[ht]
\centering
\scalebox{0.7}{
\begin{tikzpicture}
\begin{feynman}
\vertex (a){\(\nu_L\)};
\vertex[right=of a](x);
\vertex[right=of x](i);
\vertex[right=of i](j);
\vertex[right=of j](k);
\vertex[right=of k](l);
\vertex[right=of l](m);
\vertex[right=of m](n);
\vertex[right=of n](o);
\vertex[right=of o](p);
\vertex[right=of p](y);
\vertex[right=of y](b){\(\overline{\nu_L}\)};
\vertex[below=of m](d){\(H\)};
\vertex[above=of l](e){\(H\)};
\diagram*[small, horizontal=a to b, inline=(a)]{
(a)--[fermion](x)--[fermion, edge label'=\(\chi_1\)](i)--[fermion, edge label'=\(\chi_2\)](j)--[fermion, edge label'=\(\chi_3\)](k)--[fermion, edge label'=\(\chi_4\)](l)--[majorana, edge label'=\(\Xi\), insertion=0.5](m)--[anti fermion, edge label'=\(\overline{\chi_4}\)](n)--[anti fermion, edge label'=\(\overline{\chi_3}\)](o)--[anti fermion, edge label'=\(\overline{\chi_2}\)](p)--[anti fermion, edge label'=\(\overline{\chi_1}\)](y)--[anti fermion](b),
(x)--[scalar, half left, edge label=\(H\)](p),
(j)--[scalar, half left, edge label=\(H\)](n),
(y)--[scalar, half left, edge label=\(H\)](i),
(o)--[scalar, half left, edge label=\(H\)](k),
(m)--[scalar,  insertion=0.99](d),
(l)--[scalar,  insertion=0.99](e),};
\end{feynman}
\end{tikzpicture}
}
\caption{``Chain'' diagram for neutrino mass based on a model with the new fermions $\Xi \sim (1,7,0)$, $\chi_1 \sim (1, 3, -2)$, $\chi_2 \sim(1, 4, -1)$, $\chi_3 \sim (1, 5,-2)$, and $\chi_4 \sim (1,6,-1)$. This is an example for a mechanism of systematic loop suppression of the neutrino mass.}
\label{fig:rainbow}
\end{figure}
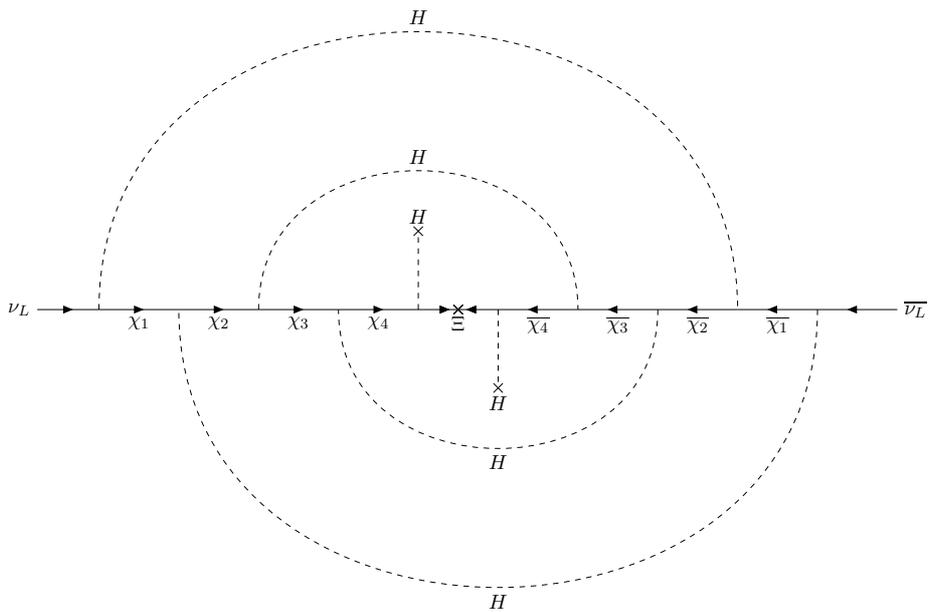

For example, for a new Majorana fermion $\Xi \sim (1,7,0)$ we need to introduce four additional new fermions. We always start with $\chi_1 \sim (1, 3, -2)$, the rest can be chosen accordingly. The model with minimal overall hypercharge would contain  $\chi_2 \sim(1, 4, -1)$,  $\chi_3 \sim (1,5,-2)$, and $\chi_4 \sim (1,6,-1)$. This will lead to a four-loop mass diagram, see Fig.~\ref{fig:rainbow}.\\

\section{Conclusions}
\label{sec:conc}

In this paper, we systematically studied radiative neutrino mass models, where the dimension-5 Weinberg operator is only generated at loop level. We add just two new beyond the Standard Model representations without employing new symmetries. The complete lists of known and new models can be found in the Tables~\ref{tab:A},~\ref{tab:BC},~\ref{tab:cA},~\ref{tab:cB}, and~\ref{tab:cC}. Thereby, two new representations is the minimal number of new fields which have to be added to the Standard Model to only generate neutrino masses at the quantum level -- with the only exception being the Zee-model~\cite{Zee1980} which just requires one new scalar representation and an additional Standard Model Higgs copy. This is a bottom-up approach to systematically study possible neutrino mass mechanisms. When considering ultraviolet completions, models which seem to be minimal from the low energy perspective can actually be non-minimal and vice versa. The study of minimal radiative neutrino masses from an ultraviolet perspective is therefore a complementary approach that might reveal a different list of minimal models.

The models C1 and C2 also contain stable neutral particles. These particles are potential dark matter candidates. The study of the interplay of neutrino and dark matter phenomenology in these models was partly done in Ref.~\cite{Cai2011} but a more complete study would be desirable.

By allowing new coloured scalars to generate neutrino masses, also scalar leptoquarks were introduced as Standard Model extensions. Scalar leptoquarks such as $\phi_{(3,2,1/3)}$, $\phi_{(3,1, -2/3)}$, and $\phi_{(3,3, -2/3)}$ are considered as possible explanations for the current B-anomalies. This opens up the window to study a common origin of neutrino masses and B-anomalies, as was already addressed by Ref.~\cite{Pas2015, Cheung2016}.

In section~\ref{sec:minFerm}, we gave a formal proof that neutrino masses cannot be generated solely via quantum effects with just two new fermionic representations. We then introduced the minimal and unique model with three new fermionic representations $\psi^{(3,-2)}_L + \psi^{(3,-2)}_R$, $\psi^{(4,-1)}_L+\psi^{(4,-1)}_R$, and $\psi^{(5,0)}_R$. As the lightest new fermion decays very weakly, this model also has a potential dark matter candidate. Furthermore, extensions of this model allow to explain large mass separations via a fermionic ladder structure, as can be inferred from the chain diagram in Fig.~\ref{fig:rainbow}. A thorough phenomenological study of the minimal fermionic Standard Model extensions which generates neutrino masses only at the quantum level is planned for future work.
 
\section*{Acknowledgements}
We thank Martin Hirsch for useful comments on the manuscript, and Evgeny Akhmedov for clarifying discussions.

\appendix
\section{Proof of uniqueness of minimal three fermion model}
\label{app:proof} 
In this appendix, we show that there is exactly one model with three new fermionic representations generating the dimension-5 Weinberg operator only at loop level without employing any new symmetries.

Let us first make some general observations for fermionic fields fulfilling 
\begin{equation}
\label{eq:cond}
\vert Y(\psi_x)\vert \geq \vert Y(H)\vert \, .
\end{equation}
For any two fermions fulfilling this condition, all Yukawa  couplings are of the form
\begin{align}
\label{eq:interactions}
&\overline{\psi_x}\Phi \psi_y : \, \, \, \mathrm{sign}\left(Y(\psi_x)\right)=\mathrm{sign}\left(Y(\psi_y)\right) \,,\\ \nonumber
&\overline{\psi_x^c}\Phi \psi_y : \, \, \, \mathrm{sign}\left(Y(\psi_x)\right)\neq\mathrm{sign}\left(Y(\psi_y)\right) \, ,
\end{align}
where $\Phi$ can be $H$ or $\tilde{H}$, but $\Phi$ could also be a mass insertion. Eq.~(\ref{eq:interactions}) can be inferred from requiring that the sum of hypercharges is zero for each term in the Lagrangian.
Thus, for any two fermion fields fulfilling eq.~(\ref{eq:cond}), the condition for lepton number conservation reads
\begin{equation}
\label{eq:esys}
\mathrm{sign}\left(Y(\psi_x)\right)L_x-\mathrm{sign}\left(Y(\psi_y)\right)L_y = 0 \, .
\end{equation}
All leptons in the Standard Model have negative hypercharge and fulfil condition~(\ref{eq:cond}). So the coupling of any new fermion obeying~(\ref{eq:cond}) to the Standard Model leptons will be of the form
\begin{align}
&\overline{X}\Phi \psi_x : \, \, \, Y(\psi_x)<0 \,,\nonumber \\
&\overline{X^c}\Phi \psi_x : \, \, \, Y(\psi_x) >0 \, ,
\end{align}
where X is any Standard Model lepton.
Eq.~(\ref{eq:SMcoupl}) tells us that all new fermion couplings to the Standard Model that are admitted satisfy eq.~(\ref{eq:cond}).
Hence, for any new fermion coupling to Standard Model leptons the equation for the lepton number is
\begin{equation}
\label{eq:hc}
L_x = -\mathrm{sign}\left(Y(\psi_x)\right)\, .
\end{equation}
Moreover, according to eq.~(\ref{eq:candis}), all fermions that the ones from eq.~(\ref{eq:SMcoupl}) can couple to obey eq.~(\ref{eq:cond}). 

Now, we consider all possible models for three new fermions, $\psi_a$, $\psi_b$, and $\psi_c$, differentiating four cases:
\begin{enumerate}
\item None of the new fermions has a direct coupling to any Standard Model leptons. In this case, one can just set $L=0$ for all new fermions and there will be no LNV. This includes the case where all new fermions have a non-trivial colour representation.
\item All of the new fields couple to the Standard Model leptons. Then all of the new fields are from the list in eq.~(\ref{eq:SMcoupl}), and obey eq.~(\ref{eq:cond}).  Considering all possible interactions of the new fermions with each other, we find, according to eq.~(\ref{eq:esys}) and eq.~(\ref{eq:hc}),
\begin{align}
\begin{split}
\mathrm{sign}\left(Y(\psi_a)\right)L_a-\mathrm{sign}\left(Y(\psi_b)\right)L_b &=-1-(-1)=0 \,,\\
\mathrm{sign}\left(Y(\psi_a)\right)L_a-\mathrm{sign}\left(Y(\psi_c)\right)L_c &=-1-(-1)=0 \,,\\
\mathrm{sign}\left(Y(\psi_c)\right)L_c-\mathrm{sign}\left(Y(\psi_b)\right)L_b &=-1-(-1)=0\, .
\end{split}
\end{align}
Hence, there is no LNV in this case.
\item Two new fermions, say $\psi_a$ and $\psi_b$ w.l.o.g., couple to the Standard Model leptons directly, the third new fermion does not. If $\psi_c$ couples to neither $\psi_a$ nor $\psi_b$, then we can just set $L_c=0$, and this reduces to the two-fermion case which has no solution. Hence, $\psi_c$ must couple to either $\psi_a$ or $\psi_b$.  In this case, it follows from eq.~(\ref{eq:esys}) and eq.~(\ref{eq:hc})
\begin{align}
\begin{split}
\mathrm{sign}\left(Y(\psi_a)\right)L_a-\mathrm{sign}\left(Y(\psi_b)\right)L_b &=-1-(-1)=0 \,,\\
\mathrm{sign}\left(Y(\psi_a)\right)L_a-\mathrm{sign}\left(Y(\psi_c)\right)L_c &=-1-\mathrm{sign}\left(Y(\psi_c)\right)L_c =0 \,,\\
\mathrm{sign}\left(Y(\psi_b)\right)L_b-\mathrm{sign}\left(Y(\psi_c)\right)L_c &=-1-\mathrm{sign}\left(Y(\psi_c)\right)L_c =0 \,.
\end{split}
\end{align}
This is solved by 
\begin{equation}
L_c = -\mathrm{sign}\left(Y(\psi_c)\right)\, .
\end{equation}
Hence, there is also no LNV in this case.
\item One field, $\psi_a$ w.l.o.g., couples directly to Standard Model leptons, the other two do not. If one of them has no coupling to the other two at all, then the problem reduces to the two fermion case and there is no LNV.

If $\psi_b$ and $\psi_c$ both have non-zero hypercharge, we find that LNV can always be avoided by choosing the lepton numbers according to eq.~(\ref{eq:hc}):
\begin{align}
\begin{split}
\mathrm{sign}\left(Y(\psi_a)\right)L_a-\mathrm{sign}\left(Y(\psi_b)\right)L_b &=-1-\mathrm{sign}\left(Y(\psi_b)\right)L_b=0\,,\\
\mathrm{sign}\left(Y(\psi_a)\right)L_a-\mathrm{sign}\left(Y(\psi_c)\right)L_c &=-1-\mathrm{sign}\left(Y(\psi_c)\right)L_c =0\,,\\
\mathrm{sign}\left(Y(\psi_b)\right)L_b-\mathrm{sign}\left(Y(\psi_c)\right)L_c &=0\,.
\end{split}
\end{align}
This is solved by
\begin{align}
\begin{split}
\mathrm{sign}\left(Y(\psi_b)\right)L_b &=-1 = \mathrm{sign}\left(Y(\psi_c)\right)L_c\\
\iff L_b = \frac{-1}{\mathrm{sign}\left(Y(\psi_b)\right)} = -\mathrm{sign}\left(Y(\psi_b)\right) &; \, \, \, \, \,  L_c =- \mathrm{sign}\left(Y(\psi_c)\right) \,.
\end{split}
\end{align}
Considering all candidates listed in eq.~(\ref{eq:candis}) and all multiplets they can couple to,
\begin{align}
 \psi^{(1,-4)}H &\sim (2,-3)\,, & \psi^{(1,-4)}\tilde{H} &\sim (2,-5)\,,\\ \nonumber
 \psi^{(3,-4)}H &\sim (2,-3)\oplus (4,-3)\,, & \psi^{(3,-4)}\tilde{H} &\sim (2,-5) \oplus (4,-5)\,,\\ \nonumber
 \psi^{(4,-1)}H&\sim (3,0) \oplus (5,0)\,, & \psi^{(4,-1)}\tilde{H} &\sim (3,-2) \oplus (5,-2)\,,\\ \nonumber
 \psi^{(4,-3)}H&\sim (3,-2) \oplus (5,-2)\,, & \psi^{(4,-3)}\tilde{H} &\sim (3,-4) \oplus (5,-4)\, ,
\end{align} 
we find that there is exactly one choice of $\psi_a$, $\psi_b$, and $\psi_c$ containing a field of zero hypercharge, such that the above argumentation does not apply. In this case,
\begin{equation}
\mathcal{L}\supset M\overline {\psi^c}_c\psi_c \, \, \Rightarrow \, \, L_c = 0\, .
\end{equation}
Since $L_a$ is determined by eq.~(\ref{eq:hc}), we find for $\psi_b$
\begin{align}
-1-\mathrm{sign}\left(Y(\psi_b)\right)L_b &=0  & \Rightarrow L_b &= -\mathrm{sign}\left(Y(\psi_b)\right)\neq 0 \,,\\ \nonumber
0\pm \mathrm{sign}\left(Y(\psi_b)\right)L_b &=0  & \Rightarrow L_b &= 0 \, .
\end{align}
\end{enumerate}
It follows that $\psi_b$ needs to couple to both $\psi_a$ and $\psi_c$ to produce lepton number violation.
We conclude that there is exaclty one model with three new fermionic multiplets and LNV namely 
\begin{align}
\psi_a &\sim (3,-2) & \psi_b &\sim (4,-1) & \psi_c &\sim (5,0) \,,
\end{align}
where the signs of all hypercharges may also be inverted. 
 
\bibliography{MEMO1}

\begin{thebibliography}{10}

\bibitem{Minkowski1977}
P.~Minkowski, ``{$\mu \to e\gamma$ at a Rate of One Out of $10^{9}$ Muon
  Decays?},'' {\em Phys. Lett.}, vol.~67B, pp.~421--428, 1977.

\bibitem{Mohapatra1979}
R.~N. Mohapatra and G.~Senjanovic, ``{Neutrino Mass and Spontaneous Parity
  Violation},'' {\em Phys. Rev. Lett.}, vol.~44, p.~912, 1980.
\newblock [,231(1979)].

\bibitem{Yanagida1979}
T.~Yanagida, ``{HORIZONTAL SYMMETRY AND MASSES OF NEUTRINOS},'' {\em Conf.
  Proc.}, vol.~C7902131, pp.~95--99, 1979.

\bibitem{GellMann1980}
M.~Gell-Mann, P.~Ramond, and R.~Slansky, ``{Complex Spinors and Unified
  Theories},'' {\em Conf. Proc.}, vol.~C790927, pp.~315--321, 1979, 1306.4669.

\bibitem{Schechter1980}
J.~Schechter and J.~W.~F. Valle, ``{Neutrino Masses in SU(2) x U(1)
  Theories},'' {\em Phys. Rev.}, vol.~D22, p.~2227, 1980.

\bibitem{Schechter1981}
J.~Schechter and J.~W.~F. Valle, ``{Neutrino Decay and Spontaneous Violation of
  Lepton Number},'' {\em Phys. Rev.}, vol.~D25, p.~774, 1982.

\bibitem{Lazarides1980}
G.~Lazarides, Q.~Shafi, and C.~Wetterich, ``{Proton Lifetime and Fermion Masses
  in an SO(10) Model},'' {\em Nucl. Phys.}, vol.~B181, pp.~287--300, 1981.

\bibitem{Mohapatra1980}
R.~N. Mohapatra and G.~Senjanovic, ``{Neutrino Masses and Mixings in Gauge
  Models with Spontaneous Parity Violation},'' {\em Phys. Rev.}, vol.~D23,
  p.~165, 1981.

\bibitem{Foot1988}
R.~Foot, H.~Lew, X.~G. He, and G.~C. Joshi, ``{Seesaw Neutrino Masses Induced
  by a Triplet of Leptons},'' {\em Z. Phys.}, vol.~C44, p.~441, 1989.

\bibitem{Vissani1997}
F.~Vissani, ``{Do experiments suggest a hierarchy problem?},'' {\em Phys.
  Rev.}, vol.~D57, pp.~7027--7030, 1998, hep-ph/9709409.

\bibitem{Casas2004}
J.~A. Casas, J.~R. Espinosa, and I.~Hidalgo, ``{Implications for new physics
  from fine-tuning arguments. 1. Application to SUSY and seesaw cases},'' {\em
  JHEP}, vol.~11, p.~057, 2004, hep-ph/0410298.

\bibitem{Abada2007}
A.~Abada, C.~Biggio, F.~Bonnet, M.~B. Gavela, and T.~Hambye, ``{Low energy
  effects of neutrino masses},'' {\em JHEP}, vol.~12, p.~061, 2007, 0707.4058.

\bibitem{Farina2013}
M.~Farina, D.~Pappadopulo, and A.~Strumia, ``{A modified naturalness principle
  and its experimental tests},'' {\em JHEP}, vol.~08, p.~022, 2013, 1303.7244.

\bibitem{Clarke2015}
J.~D. Clarke, R.~Foot, and R.~R. Volkas, ``{Electroweak naturalness in the
  three-flavor type I seesaw model and implications for leptogenesis},'' {\em
  Phys. Rev.}, vol.~D91, no.~7, p.~073009, 2015, 1502.01352.

\bibitem{Fabbrichesi2015}
M.~Fabbrichesi and A.~Urbano, ``{Naturalness redux: The case of the neutrino
  seesaw mechanism},'' {\em Phys. Rev.}, vol.~D92, p.~015028, 2015, 1504.05403.

\bibitem{Clarke2015b}
J.~D. Clarke, R.~Foot, and R.~R. Volkas, ``{Natural leptogenesis and neutrino
  masses with two Higgs doublets},'' {\em Phys. Rev.}, vol.~D92, no.~3,
  p.~033006, 2015, 1505.05744.

\bibitem{Chabab2015}
M.~Chabab, M.~C. Peyranère, and L.~Rahili, ``{Naturalness in a type II seesaw
  model and implications for physical scalars},'' {\em Phys. Rev.}, vol.~D93,
  no.~11, p.~115021, 2016, 1512.07280.

\bibitem{Clarke2016}
J.~D. Clarke and P.~Cox, ``{Naturalness made easy: two-loop naturalness bounds
  on minimal SM extensions},'' {\em JHEP}, vol.~02, p.~129, 2017, 1607.07446.

\bibitem{Salvio2016}
A.~Salvio, ``{Solving the Standard Model Problems in Softened Gravity},'' {\em
  Phys. Rev.}, vol.~D94, no.~9, p.~096007, 2016, 1608.01194.

\bibitem{Bambhaniya2016}
G.~Bambhaniya, P.~Bhupal~Dev, S.~Goswami, S.~Khan, and W.~Rodejohann,
  ``{Naturalness, Vacuum Stability and Leptogenesis in the Minimal Seesaw
  Model},'' {\em Phys. Rev.}, vol.~D95, no.~9, p.~095016, 2017, 1611.03827.

\bibitem{Dev2017}
P.~S.~B. Dev, C.~M. Vila, and W.~Rodejohann, ``{Naturalness in testable type II
  seesaw scenarios},'' {\em Nucl. Phys.}, vol.~B921, pp.~436--453, 2017,
  1703.00828.

\bibitem{Zee1980}
A.~Zee, ``{A Theory of Lepton Number Violation, Neutrino Majorana Mass, and
  Oscillation},'' {\em Phys. Lett.}, vol.~93B, p.~389, 1980.
\newblock [Erratum: Phys. Lett.95B,461(1980)].

\bibitem{Zee1985}
A.~Zee, ``{Quantum Numbers of Majorana Neutrino Masses},'' {\em Nucl. Phys.},
  vol.~B264, pp.~99--110, 1986.

\bibitem{Babu1988}
K.~S. Babu, ``{Model of 'Calculable' Majorana Neutrino Masses},'' {\em Phys.
  Lett.}, vol.~B203, pp.~132--136, 1988.

\bibitem{Perez2009}
P.~Fileviez~Perez and M.~B. Wise, ``{On the Origin of Neutrino Masses},'' {\em
  Phys. Rev.}, vol.~D80, p.~053006, 2009, 0906.2950v2.

\bibitem{Ma2006}
E.~Ma, ``{Verifiable radiative seesaw mechanism of neutrino mass and dark
  matter},'' {\em Phys. Rev.}, vol.~D73, p.~077301, 2006, hep-ph/0601225.

\bibitem{Babu2001}
K.~S. Babu and C.~N. Leung, ``{Classification of effective neutrino mass
  operators},'' {\em Nucl. Phys.}, vol.~B619, pp.~667--689, 2001,
  hep-ph/0106054v1.

\bibitem{Angel2012}
P.~W. Angel, N.~L. Rodd, and R.~R. Volkas, ``{Origin of neutrino masses at the
  LHC: $\Delta L = 2$ effective operators and their ultraviolet completions},''
  {\em Phys. Rev.}, vol.~D87, no.~7, p.~073007, 2013, 1212.6111v2.

\bibitem{Bonnet2012}
F.~Bonnet, M.~Hirsch, T.~Ota, and W.~Winter, ``{Systematic study of the d=5
  Weinberg operator at one-loop order},'' {\em JHEP}, vol.~07, p.~153, 2012,
  1204.5862v2.

\bibitem{Sierra2014}
D.~Aristizabal~Sierra, A.~Degee, L.~Dorame, and M.~Hirsch, ``{Systematic
  classification of two-loop realizations of the Weinberg operator},'' {\em
  JHEP}, vol.~03, p.~040, 2015, 1411.7038.

\bibitem{Cai2014}
Y.~Cai, J.~D. Clarke, M.~A. Schmidt, and R.~R. Volkas, ``{Testing Radiative
  Neutrino Mass Models at the LHC},'' {\em JHEP}, vol.~02, p.~161, 2015,
  1410.0689v3.

\bibitem{Cai2017}
Y.~Cai, J.~Herrero-García, M.~A. Schmidt, A.~Vicente, and R.~R. Volkas,
  ``{From the trees to the forest: a review of radiative neutrino mass
  models},'' {\em Front.in Phys.}, vol.~5, p.~63, 2017, 1706.08524.

\bibitem{Cepedello2017}
R.~Cepedello, M.~Hirsch, and J.~C. Helo, ``{Loop neutrino masses from $d = 7$
  operator},'' {\em JHEP}, vol.~07, p.~079, 2017, 1705.01489.

\bibitem{Ma2009}
E.~Ma, ``{Neutrino Mass: Mechanisms and Models},'' 2009, 0905.0221v1.

\bibitem{Ma1998}
E.~Ma, ``{Pathways to naturally small neutrino masses},'' {\em Phys. Rev.
  Lett.}, vol.~81, pp.~1171--1174, 1998, hep-ph/9805219v4.

\bibitem{Sierra2015}
D.~Aristizabal~Sierra, ``{Two-loop-induced neutrino masses: A model-independent
  perspective},'' {\em PoS}, vol.~PLANCK2015, p.~008, 2015, 1510.04958v1.

\bibitem{Anamiati2018}
G.~Anamiati, O.~Castillo-Felisola, R.~M. Fonseca, J.~C. Helo, and M.~Hirsch,
  ``{High-dimensional neutrino masses},'' {\em JHEP}, vol.~12, p.~066, 2018,
  1806.07264.

\bibitem{Law2013}
S.~S.~C. Law and K.~L. McDonald, ``{The simplest models of radiative neutrino
  mass},'' {\em Int. J. Mod. Phys.}, vol.~A29, p.~1450064, 2014, 1303.6384v1.

\bibitem{Bonilla2018}
C.~Bonilla, S.~Centelles-Chuliá, R.~Cepedello, E.~Peinado, and R.~Srivastava,
  ``{Dark matter stability and Dirac neutrinos using only Standard Model
  symmetries},'' 2018, 1812.01599.

\bibitem{Calle2018}
J.~Calle, D.~Restrepo, C.~E. Yaguna, and s.~Zapata, ``{Minimal radiative Dirac
  neutrino mass models},'' 2018, 1812.05523.

\bibitem{Chulia2018}
S.~Centelles~Chuliá, R.~Srivastava, and J.~W.~F. Valle, ``{Seesaw roadmap to
  neutrino mass and dark matter},'' {\em Phys. Lett.}, vol.~B781, pp.~122--128,
  2018, 1802.05722.

\bibitem{Chulia2018a}
S.~Centelles~Chuliá, R.~Srivastava, and J.~W.~F. Valle, ``{Seesaw Dirac
  neutrino mass through dimension-six operators},'' {\em Phys. Rev.}, vol.~D98,
  no.~3, p.~035009, 2018, 1804.03181.

\bibitem{Babu2009}
K.~S. Babu, S.~Nandi, and Z.~Tavartkiladze, ``{New Mechanism for Neutrino Mass
  Generation and Triply Charged Higgs Bosons at the LHC},'' {\em Phys. Rev.},
  vol.~D80, p.~071702, 2009, 0905.2710v2.

\bibitem{Liao2010}
Y.~Liao, ``{Cascade Seesaw for Tiny Neutrino Mass},'' {\em JHEP}, vol.~06,
  p.~098, 2011, 1011.3633v3.

\bibitem{Kumericki2012}
K.~Kumericki, I.~Picek, and B.~Radovcic, ``{TeV-scale Seesaw with Quintuplet
  Fermions},'' {\em Phys. Rev.}, vol.~D86, p.~013006, 2012, 1204.6599.

\bibitem{Picek2012}
I.~Picek and B.~Radovcic, ``{Enhancement of $h \to \gamma \gamma$ by
  seesaw-motivated exotic scalars},'' {\em Phys. Lett.}, vol.~B719,
  pp.~404--408, 2013, 1210.6449.

\bibitem{McDonald2013}
K.~L. McDonald, ``{Minimal Tree-Level Seesaws with a Heavy Intermediate
  Fermion},'' {\em JHEP}, vol.~07, p.~020, 2013, 1303.4573.

\bibitem{Chen2013}
C.-S. Chen and Y.-J. Zheng, ``{LHC signatures for the cascade seesaw
  mechanism},'' {\em PTEP}, vol.~2015, p.~103B02, 2015, 1312.7207.

\bibitem{Ding2014}
R.~Ding, Z.-L. Han, Y.~Liao, H.-J. Liu, and J.-Y. Liu, ``{Phenomenology in the
  minimal cascade seesaw mechanism for neutrino masses},'' {\em Phys. Rev.},
  vol.~D89, no.~11, p.~115024, 2014, 1403.2040.

\bibitem{Cai2011}
Y.~Cai, X.-G. He, M.~Ramsey-Musolf, and L.-H. Tsai, ``{R$\nu$MDM and Lepton
  Flavor Violation},'' {\em JHEP}, vol.~12, p.~054, 2011, 1108.0969v4.

\bibitem{Babu2010}
K.~S. Babu and J.~Julio, ``{Two-Loop Neutrino Mass Generation through
  Leptoquarks},'' {\em Nucl. Phys.}, vol.~B841, pp.~130--156, 2010, 1006.1092.

\bibitem{Dorsner2017}
I.~Doršner, S.~Fajfer, and N.~Košnik, ``{Leptoquark mechanism of neutrino
  masses within the grand unification framework},'' {\em Eur. Phys. J.},
  vol.~C77, no.~6, p.~417, 2017, 1701.08322.

\bibitem{Pas2015}
H.~Päs and E.~Schumacher, ``{Common origin of $R_K$ and neutrino masses},''
  {\em Phys. Rev.}, vol.~D92, no.~11, p.~114025, 2015, 1510.08757.

\bibitem{Cheung2016}
K.~Cheung, T.~Nomura, and H.~Okada, ``{Testable radiative neutrino mass model
  without additional symmetries and explanation for the $b \to s \ell^+ \ell^-$
  anomaly},'' {\em Phys. Rev.}, vol.~D94, no.~11, p.~115024, 2016, 1610.02322.

\bibitem{Chang2016}
W.-F. Chang, S.-C. Liou, C.-F. Wong, and F.~Xu, ``{Charged Lepton Flavor
  Violating Processes and Scalar Leptoquark Decay Branching Ratios in the
  Colored Zee-Babu Model},'' {\em JHEP}, vol.~10, p.~106, 2016, 1608.05511.

\bibitem{Aaij2014}
R.~Aaij {\em et~al.}, ``{Test of lepton universality using $B^{+}\rightarrow
  K^{+}\ell^{+}\ell^{-}$ decays},'' {\em Phys. Rev. Lett.}, vol.~113,
  p.~151601, 2014, 1406.6482.

\bibitem{Hiller2014}
G.~Hiller and M.~Schmaltz, ``{$R_K$ and future $b \to s \ell \ell$ physics
  beyond the standard model opportunities},'' {\em Phys. Rev.}, vol.~D90,
  p.~054014, 2014, 1408.1627.

\bibitem{Becirevic2016}
D.~Bečirević, S.~Fajfer, N.~Košnik, and O.~Sumensari, ``{Leptoquark model to
  explain the $B$-physics anomalies, $R_K$ and $R_D$},'' {\em Phys. Rev.},
  vol.~D94, no.~11, p.~115021, 2016, 1608.08501.

\bibitem{Becirevic2016b}
D.~Bečirević, N.~Košnik, O.~Sumensari, and R.~Zukanovich~Funchal,
  ``{Palatable Leptoquark Scenarios for Lepton Flavor Violation in Exclusive
  $b\to s\ell_1\ell_2$ modes},'' {\em JHEP}, vol.~11, p.~035, 2016, 1608.07583.

\bibitem{Lees2013}
J.~P. Lees {\em et~al.}, ``{Measurement of an Excess of $\bar{B} \to
  D^{(*)}\tau^- \bar{\nu}_\tau$ Decays and Implications for Charged Higgs
  Bosons},'' {\em Phys. Rev.}, vol.~D88, no.~7, p.~072012, 2013, 1303.0571.

\bibitem{Huschle2015}
M.~Huschle {\em et~al.}, ``{Measurement of the branching ratio of $\bar{B} \to
  D^{(\ast)} \tau^- \bar{\nu}_\tau$ relative to $\bar{B} \to D^{(\ast)} \ell^-
  \bar{\nu}_\ell$ decays with hadronic tagging at Belle},'' {\em Phys. Rev.},
  vol.~D92, no.~7, p.~072014, 2015, 1507.03233.

\bibitem{Aaij2015}
R.~Aaij {\em et~al.}, ``{Measurement of the ratio of branching fractions
  $\mathcal{B}(\bar{B}^0 \to
  D^{*+}\tau^{-}\bar{\nu}_{\tau})/\mathcal{B}(\bar{B}^0 \to
  D^{*+}\mu^{-}\bar{\nu}_{\mu})$},'' {\em Phys. Rev. Lett.}, vol.~115, no.~11,
  p.~111803, 2015, 1506.08614.
\newblock [Erratum: Phys. Rev. Lett.115,no.15,159901(2015)].

\bibitem{Fajfer2012}
S.~Fajfer, J.~F. Kamenik, and I.~Nisandzic, ``{On the $B \to D^* \tau \bar
  \nu_{\tau}$ Sensitivity to New Physics},'' {\em Phys. Rev.}, vol.~D85,
  p.~094025, 2012, 1203.2654.

\bibitem{Becirevic2012}
D.~Bečirević, N.~Košnik, and A.~Tayduganov, ``{$\bar B\to D\tau\bar
  \nu_\tau$ vs. $\bar B\to D\mu\bar \nu_\mu$},'' {\em Phys. Lett.}, vol.~B716,
  pp.~208--213, 2012, 1206.4977.

\bibitem{Bauer2015}
M.~Bauer and M.~Neubert, ``{Minimal Leptoquark Explanation for the
  R$_{D^{(*)}}$ , R$_K$ , and $(g-2)_g$ Anomalies},'' {\em Phys. Rev. Lett.},
  vol.~116, no.~14, p.~141802, 2016, 1511.01900.

\bibitem{Nath2006}
P.~Nath and P.~Fileviez~Perez, ``{Proton stability in grand unified theories,
  in strings and in branes},'' {\em Phys. Rept.}, vol.~441, pp.~191--317, 2007,
  hep-ph/0601023.

\bibitem{Hagedorn2016}
C.~Hagedorn, T.~Ohlsson, S.~Riad, and M.~A. Schmidt, ``{Unification of Gauge
  Couplings in Radiative Neutrino Mass Models},'' {\em JHEP}, vol.~09, p.~111,
  2016, 1605.03986.

\bibitem{Babu2011}
K.~S. Babu and J.~Julio, ``{Radiative Neutrino Mass Generation through
  Vector-like Quarks},'' {\em Phys. Rev.}, vol.~D85, p.~073005, 2012,
  1112.5452.

\bibitem{Popov2016}
O.~Popov and G.~A. White, ``{One Leptoquark to unify them? Neutrino masses and
  unification in the light of $(g-2)_\mu$, $R_{D^{(\star)}}$ and $R_K$
  anomalies},'' {\em Nucl. Phys.}, vol.~B923, pp.~324--338, 2017, 1611.04566.

\bibitem{Angel2013}
P.~W. Angel, Y.~Cai, N.~L. Rodd, M.~A. Schmidt, and R.~R. Volkas, ``{Testable
  two-loop radiative neutrino mass model based on an $LLQd^cQd^c$ effective
  operator},'' {\em JHEP}, vol.~10, p.~118, 2013, 1308.0463.
\newblock [Erratum: JHEP11,092(2014)].

\bibitem{Cai2017a}
Y.~Cai, J.~Gargalionis, M.~A. Schmidt, and R.~R. Volkas, ``{Reconsidering the
  One Leptoquark solution: flavor anomalies and neutrino mass},'' {\em JHEP},
  vol.~10, p.~047, 2017, 1704.05849.

\bibitem{Gross2018}
C.~Gross, A.~Mitridate, M.~Redi, A.~Strumia, and J.~Smirnov, ``{Cosmological
  Abundance of Colored Relics},'' 2018, 1811.08418.

\bibitem{DeLuca2018}
V.~De~Luca, A.~Mitridate, M.~Redi, J.~Smirnov, and A.~Strumia, ``{Colored Dark
  Matter},'' {\em Phys. Rev.}, vol.~D97, no.~11, p.~115024, 2018, 1801.01135.

\bibitem{Froggatt1978}
C.~D. Froggatt and H.~B. Nielsen, ``{Hierarchy of Quark Masses, Cabibbo Angles
  and CP Violation},'' {\em Nucl. Phys.}, vol.~B147, pp.~277--298, 1979.

\bibitem{Choi2015}
K.~Choi and S.~H. Im, ``{Realizing the relaxion from multiple axions and its UV
  completion with high scale supersymmetry},'' {\em JHEP}, vol.~01, p.~149,
  2016, 1511.00132v2.

\bibitem{Kaplan2015}
D.~E. Kaplan and R.~Rattazzi, ``{Large field excursions and approximate
  discrete symmetries from a clockwork axion},'' {\em Phys. Rev.}, vol.~D93,
  no.~8, p.~085007, 2016, 1511.01827v1.

\end{thebibliography}
\bibliographystyle{hieeetr}

\end{document}